\documentclass[times,twocolumn,final,authoryear]{elsarticle}

\usepackage{ycviu}
\usepackage{framed,multirow}

\usepackage{latexsym}

\usepackage{url}
\usepackage{xcolor}
\definecolor{newcolor}{rgb}{.8,.349,.1}

\usepackage{cite}
\usepackage{amsmath,amssymb,amsfonts}
\usepackage{algorithmic}
\usepackage{graphicx}
\usepackage{textcomp}
\usepackage{algorithm}
\usepackage{comment}
\usepackage{subcaption}
\usepackage{multirow}
\journal{Computer Vision and Image Understanding}

\begin{document}

\thispagestyle{empty}

\clearpage
\thispagestyle{empty}

\ifpreprint
  \vspace*{-1pc}
\else
\fi

\begin{table*}[!t]
\ifpreprint\else\vspace*{-15pc}\fi

This paper was published in Computer Vision and Image Understanding, Elsevier. The link to published version is as follows:
\\

https://www.sciencedirect.com/science/article/pii/S107731422300098X
\\

Please cite this paper as follows:
\\

Hasan F. Ates, Suleyman Yildirim, Bahadir K. Gunturk,
Deep learning-based blind image super-resolution with iterative kernel reconstruction and noise estimation,
Computer Vision and Image Understanding,
Volume 233,
2023,
103718,
ISSN 1077-3142,
https://doi.org/10.1016/j.cviu.2023.103718.
\\

The code of IKR-Net is avaiable in github:
\\

https://github.com/hfates/IKR-Net

\section*{Research Highlights}


\vskip1pc

\fboxsep=6pt
\fbox{
\begin{minipage}{.95\textwidth}
\vskip1pc
\begin{itemize}

\item An iterative deep network architecture is proposed for blind single-image super-resolution.
\item The model contains separate modules for image reconstruction, blur kernel estimation and noise estimation.
\item The model achieves state-of-the-art results for noisy images that contain motion blur. 

\end{itemize}
\vskip1pc
\end{minipage}
}

\end{table*}

\clearpage

\ifpreprint
  \setcounter{page}{1}
\else
  \setcounter{page}{1}
\fi

\begin{frontmatter}

\title{Deep learning-based blind image super-resolution \\with iterative kernel reconstruction and noise estimation}

\author[1]{Hasan F. \snm{Ates}\corref{cor1}} 
\cortext[cor1]{Corresponding author: 
  Tel.: +90-216-564-9855}  
\ead{hasan.ates@ozyegin.edu.tr}
\author[2]{Suleyman  \snm{Yildirim}}
\author[3]{Bahadir K. \snm{Gunturk}}

\address[1]{Faculty of Engineering, Ozyegin University, Istanbul, Turkey}
\address[2]{College of Engineering, Koc University, Istanbul, Turkey}
\address[3]{School of Engineering and Natural Sciences, Istanbul Medipol University, Istanbul, Turkey}

\received{1 May 2013}
\finalform{10 May 2013}
\accepted{13 May 2013}
\availableonline{15 May 2013}
\communicated{S. Sarkar}

\begin{abstract}
Blind single image super-resolution (SISR) is a challenging task in image processing due to the ill-posed nature of the inverse problem. Complex degradations present in real life images make it difficult to solve this problem using naïve deep learning approaches, where models are often trained on synthetically generated image pairs. Most of the effort so far has been focused on solving the inverse problem under some constraints, such as for a limited space of blur kernels and/or assuming noise-free input images. Yet, there is a gap in the literature to provide a well-generalized deep learning-based solution that performs well on images with unknown and highly complex degradations. In this paper, we propose IKR-Net (Iterative Kernel Reconstruction Network) for blind SISR. In the proposed approach, kernel and noise estimation and high-resolution image reconstruction are carried out iteratively using dedicated deep models. The iterative refinement provides significant improvement in both the reconstructed image and the estimated blur kernel even for noisy inputs. IKR-Net provides a generalized solution that can handle any type of blur and level of noise in the input low-resolution image. IKR-Net achieves state-of-the-art results in blind SISR, especially for noisy images with motion blur.  
\end{abstract}
\begin{keyword}
\KWD Super-resolution\sep Blind\sep Iterative\sep Deep network 
\end{keyword}

\end{frontmatter}

\section{Introduction}
Solving inverse problems in image and video processing has always been a challenging area for researchers in the field of deep learning. In particular, the ability to develop deep learning-based models for problems such as single image super-resolution (SISR) is promising. But still, it requires further study to apply in real world scenarios where blur models are unknown \citep{b1}, \citep{b2}. Due to many environmental factors or device limitations during photography, captured images are exposed to blurring and noise in different forms. The simple form of the forward degradation model can be described as:
\begin{equation}
    y = (x\:\otimes\:k)\downarrow_{s}\,+\;n
\label{eq1}
\end{equation}
where \emph{x} is high-resolution (HR) and \emph{y} is low-resolution (LR) image, $\downarrow_{s}$ denotes the s-fold downsampling operation (which is keeping the upper-left pixel for each distinct $\emph{s}\times \emph{s}$ patch), $\otimes$ denotes convolution of the blur kernel \emph{k} with HR image \emph{x}, and  \emph{n} is additive noise. The inverse problem, on the other hand, can be modeled as a function of \emph{y}, \emph{s}, \emph{k}, $\sigma$ (i.e. noise standard deviation) which provides a mapping from the degraded image \emph{y} to the latent original image \emph{x}.

\begin{figure*}[htbp]
\centering
\includegraphics[scale=0.50]{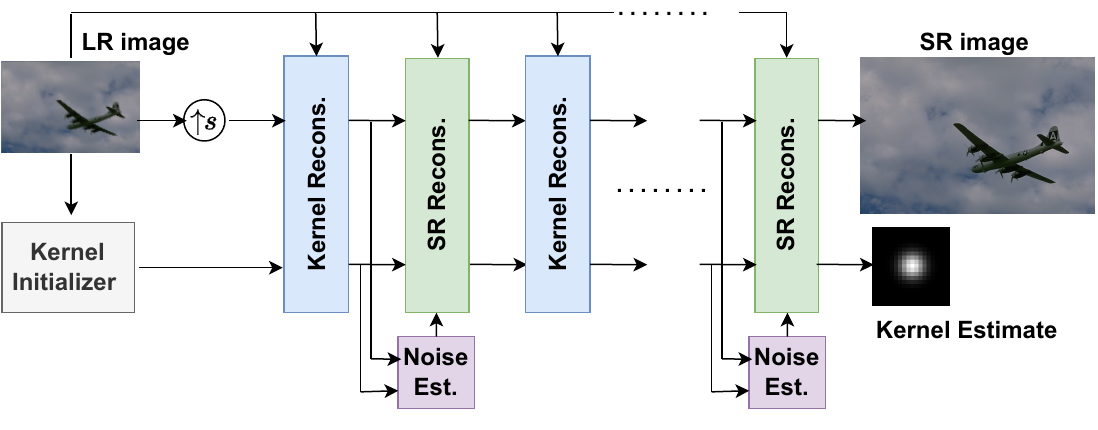} 
\caption{Iterative estimation of SR image and blur kernel in IKR-Net architecture.}
\label{fig1}
\end{figure*}

The purpose of deep learning based SISR is to find an approximate solution for this inverse function. Fundamentally, these methods are aimed at learning the inverse mapping from blurry LR to sharp HR image by training on LR-HR image pairs. Traditionally, there are two main approaches: 1) learning the inverse mapping directly from data without explicit degradation modelling; 2) modelling the degradation parameters explicitly to utilize it as prior knowledge. For the first approach, Generative Adversarial Networks (GANs) are one of the most popular tools due to its ability to learn complex statistical distributions. Most of these methods are trained on synthetically generated datasets \citep{b3, b6}, i.e. LR images are generated from HR images with some predefined blur kernels. These blur kernels are typically assumed to be known and fixed (e.g. bicubic) or belong to a limited space of Gaussian kernels. Thus, they cannot represent the real life degradations adequately and the performance of these models significantly drop when tested on noisy images filtered with complex blur kernels. In reality, images are affected by various types of blur, and since these data-driven models rely heavily on the limited dataset that they are trained on, their generalization capability is low. To overcome this obstacle, some work in literature enhances their training set by generating more realistic degradation models \citep{b9, b10, b25, b26, b29}. Although these efforts improve super-resolution performance on real-life images to a certain extent, the model's generalization capability is limited by the extent of the training dataset, which leads to poor results or loss of quality when the image is exposed to a complex degradation that is not covered by the training set.

For the second approach that models the degradation explicitly, most of the current work is focused on predicting a blur kernel representation from an input LR image and fusing the predicted representation into the SR reconstruction network \citep{b12, b13, b14, b30, b33}. Typically these methods prefer to predict the blur kernel in a low dimensional feature space, rather than to reconstruct the actual kernel. Moreover, most of the blur modelling SR solutions perform under the assumption of Gaussian kernels and overlook the more complicated motion blurs. In addition, noise modelling in the input image is generally ignored in blind methods, which substantially lowers their applicability to real-life cases. Therefore, a complete modelling of the full degradation process, including all types of blur and additive noise, is crucial in order to achieve an adaptive, generalized deep learning-based solution to SISR problem.

In this paper, an iterative deep learning solution is proposed for blind SISR, named as Iterative Kernel Reconstruction network (IKR-Net). IKR-Net combines the blindness of data-driven methods with the generalization capability of model-based methods. The iterative update steps in IKR-Net model are visualized in Figure \ref{fig1}. The detailed architecture is provided in Figure \ref{fig2}. IKR-Net consists of four distinct modules: Kernel Initializer module ($\mathcal{I}$) produces the initial kernel estimate from the input LR image, Kernel Reconstruction module ($\mathcal{D}k/\mathcal{P}k$) iteratively corrects the estimated kernel, Noise Estimator module ($\mathcal{F}$) estimates the noise standard deviation and related model parameters, and SR Reconstruction module ($\mathcal{D/P}$) iteratively reconstructs the image. With IKR-Net, we extend the work of \citet{b15} to blind SISR and further improve our previous model in \citet{b34} by incorporating noise estimation into the iterative architecture. Results show that the quality of output SR images become close to that of non-blind methods despite any prior knowledge of the blur kernel and/or noise level. IKR-Net provides state-of-the-art performance consistently for any type of blur kernel (i.e. isotropic/anisotropic Gaussian, motion blur) and for both noisy and noise-free images. IKR-Net significantly outperforms existing blind SR methods visually and in objective metrics for complex motion kernels and for noisy images. The novel contributions of this paper are as follows:

\begin{figure*}[htbp]
\centering
\includegraphics[scale=0.70]{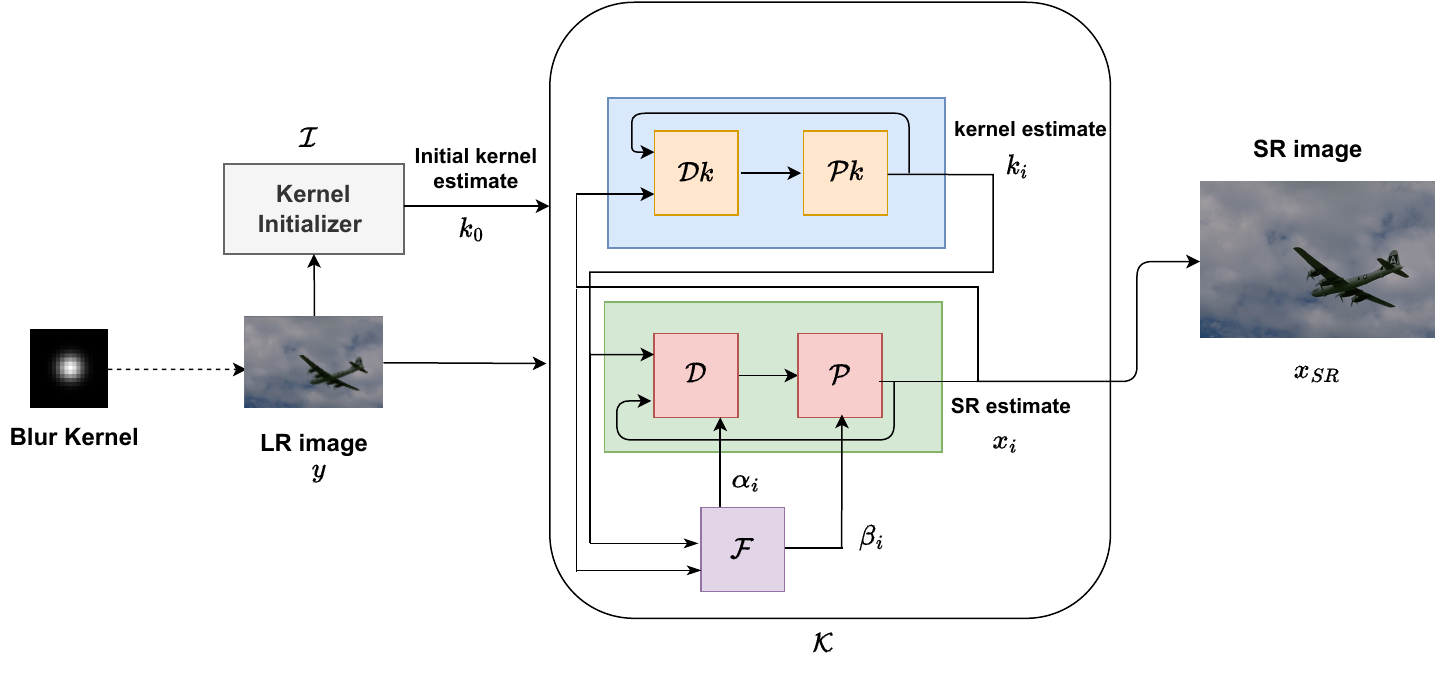} 
\caption{The overall iterative architecture of IKR-Net model.}
\label{fig2}
\end{figure*}

\begin{itemize}
\item We propose an end-to-end trainable iterative deep SISR framework, that contains multiple separate modules for kernel initialization, iterative kernel estimation, noise estimation and iterative SR image reconstruction.
\item We apply a model-based and learning based joint approach to both kernel and SR image reconstruction. Deep learning is used for regularization during iterative estimation of the kernel and the SR image.
\item We provide a well-generalized solution to blind SISR that exhibits robust performance for any complex degradation, including nonlinear motion blurs and additive noise. The enhanced model architecture proposed in this paper is superior to our previous work in \citet{b34} and \citet{b16}, and produces high-quality SR images even in the presence of noise. 
\item
The proposed model is tested with 12 selected isotropic/ anisotropic Gaussian and motion blur kernels \citep{b15} using multiple benchmark datasets with and without noise. We achieve state-of-the-art results in blind SISR, and approach the quality level of non-blind methods.
\end{itemize}

When compared to our previous work in \citet{b34}, the super-resolution model in this paper contains noise and hyper-parameter estimation modules in order to handle noise in input LR images. As a result, a more generalized model is obtained that shows superior performance not only for synthetically generated but also for real-life LR images. In addition, this paper contains extended simulation results for scales $\times 4$ and $\times 2$ with noisy and noise-free images and comparisons with six state-of-the-art methods from literature.  Visual comparisons for real blurry images are provided as well. This paper also contains an ablation study, in which several features and parameters of the proposed architecture are further analyzed, with special focus on the SR performance gains obtained by the iterative noise estimation and removal steps.

In this paper, we propose a generic solution to blind SISR problem by combining learning-based and model-based approaches in an end-to-end trainable iterative network. Similar to model-based methods, IKR-Net can effectively handle complex degradations with different types of blur kernels at various scale factors and input noise levels. The iterative and joint refinement of both the SR and blur kernel estimates makes it possible to achieve superior results under various degradation settings. The modules used for initial kernel estimation, noise estimation, SR image and kernel reconstruction are generic convolutional neural networks. These modules are initially trained independently for their respective tasks and then put together and fine-tuned in an iterative and end-to-end manner, as depicted in Figure \ref{fig2}. This plug-and-play nature of the proposed architecture makes it possible to replace any of the given modules with alternative solutions and/or train the model effectively and efficiently for any degradation setting.

Section 2 discusses the previous work in deep learning-based blind SISR. Section 3 gives the architectural details of IKR-Net. Section 4 provides details of training, simulations and evaluation of results. Section 5 concludes the paper with discussion of future work.

\section{Related Work}
Learning-based SISR models usually require LR-HR pairs dataset to train a model. Typically LR images are generated from HR images with some predefined blur kernels. Mostly, the bicubic blur kernel is used to downsample the HR images to generate the LR-HR pair for SR model training. Multiple deep learning models \citep{b21+, b22+, b23+, b24+, b25+, b26+} have been used for the super resolution of the image using bicubic downsampling. Assuming a fixed blur kernel in these models leads to decreased performance when applied in real world images, because the real world blur kernel is different and complex from the assumed kernels. These models can perform well with the bicubic degradation setting but cannot be used directly to upsample real world images.

The traditional blind SR approach \citep{b6+, b7+} uses model-based optimization. \citet{b8+} use comparable patches to look for similar patterns and estimate the blur kernel. However, if there is noise in the image, the estimation accuracy of these models will be greatly reduced. ZSSR \citep{b11+} is trained on a single image to take full advantage of its unique inside information. The model is trained at various scales of the input image and utilizes the kernel estimation algorithms of \citet{b8+} when the blur kernel is unknown. Traditional model-based algorithms use various regularization techniques in order to achieve accurate and artifact-free blur kernel and SR image reconstruction.  

Data-driven approaches in SISR rely heavily on the content of the training set for optimizing SR performance. Model-based techniques, on the other hand, typically assume fixed apriori models to estimate the blur kernel and reconstruct the SR image. In the following sections we review the existing literature in both directions  and explain our novel framework to bring the two approaches together in an learning-based iterative optimization setting.


\subsection{Data Driven Approach}
Data driven SISR methods aim to learn the inverse mapping from degraded LR images to clean HR images by training the deep SR model directly on LR-HR image pairs in an end-to-end manner. There are numerous deep learning models proposed to learn this mapping \citep{b9, b10, b24, b25, b26, b27, b28, b29}. \citet{b3} propose SRGAN as a solution to the problem of classical SISR. The model uses a pre-trained feature extraction network to improve perceptual quality and a loss function to eliminate contextual errors between feature maps instead of low-level pixel-based error measurements. Consequently, the output image is prevented from being over-smoothed to achieve a high signal-to-noise ratio (SNR). Following SRGAN, \citet{b4} developed ESRGAN to further improve visual quality. In this method, SRGAN’s network structure is modified as well as its perceptual loss function. Batch Normalization layers are removed from the architecture and Residual in Residual Dense Block structure is used as the basic block in the generator network. In addition, the discriminator network is improved using the Relativistic GAN method. The new discriminator estimates not only the probability that an image is real or fake, but also the probability that one is more realistic than the other. Over the next few years, various GAN-based approaches have been introduced to improve the perceptual quality.  \citet{b5} propose a ranker network model (RankSRGAN) that can learn these perceptual metrics that are highly dependent on human vision. \citet{b6} calculate the reconstruction difficulty for each pixel by obtaining a score map, rather than obtaining a single discriminator score for each image. Additionally, instead of using two completely separate structures, a common feature extraction network is used for the generator and discriminator networks. However, as already mentioned, these methods train on synthetically degraded images using only predefined kernels which are mostly bicubic and/or Gaussian. Thus, they rely heavily on the limited dataset they are trained on and their generalization capability is inadequate for real-world. Consequently, performance is far from being reliable when tested on realistic images exposed to different types of blurring and noise.

Many researchers have tried to develop a solution that can work on realistic images with various forward degradation models. Some existing work learns the statistical domain distribution of natural source images using an unsupervised learning approach \citep{b24, b28}, which trains the network with real images only. Others have tried to create realistic datasets that take more complex degradation models into account \citep{b9, b10, b25, b26, b29}.  \citet{b9} provides an approach to blur modeling and uses blur models extracted from natural images to prepare realistic datasets. They use the dark channel \citep{b11} to estimate realistic blur kernels from real LR images. Then, a GAN network is trained with these estimated blur kernels to create a large realistic blur model pool. A kernel is randomly selected from the kernel pool and the LR image is generated for the training set. \citet{b10} propose a similar approach using realistic degradation model pool generated by estimating blur and noise in natural images. However, model performance relies heavily on the kernel estimation algorithm \citep{b11}. \citet{b25} propose utilizing a second order degradation process to generate more practical LR images by applying the classical degradation model twice with the addition of JPEG compression. Another practical degradation model is proposed by \citet{b26}. They basically create multiple sets of parameters for each degradation type (e.g., blurring, resizing, noise) and randomly sample a number of degradation operations to obtain diverse degradation sequences. These sequences of operations are applied to HR images to generate corresponding LR images.  \citet{b27} trains a Swin Transformer architecture with the degradation process proposed by \citet{b26} to achieve a generalized solution for real-world SR. \citet{b29} aims to learn the source domain distribution corruptions to generate realistic LR-HR image pairs.  Unfortunately, these methods are not sufficient to provide robust performance for all real-world images, because the degradation models used in training are not comprehensive enough to handle all types of degradations.

\subsection{Explicit Degradation Modelling Approach}
These methods generally aim to estimate the degradation model first or define some specific priors in order to perform successfully regardless of the degradation model \citep{b12, b13, b14, b15, b16, b30, b31, b32, b33}. In the method proposed by \citet{b12}, blur model is estimated by using the internal distribution similarity between the patch cropped directly from the LR image and the patch cropped from the re-blurred and downscaled version of the same image. \citet{b13} and \citet{b14} iterate through the SR process by first estimating the blur model in a low dimensional space and then integrating the features of encoded blur model into their reconstruction network. Intermediate kernel representation and SR image outputs are corrected at each step iteratively. \citet{b30} propose an unsupervised method to represent degradations in a low dimensional space using contrastive learning. Then, learned representations are fed into a degradation-aware SR network applying feature adaptation and channel modulation. \citet{b33} used reinforcement learning to estimate the blur kernels through a non-differentiable perceptual metric. However, these methods generally restrict the search space of blur kernels in one way or the other and and they estimate the kernel in this reduced dimensional subspace. \citet{b32} perform spatially-varying kernel estimation and propose adjusting intermediate image features based on the predicted degradation kernels. \citet{b35} propose Flow Kernel Prior (FKP), a kernel estimation method, based on normalizing flow that allows to deform the complex data distribution on to a simple and tractable distribution. \citet{cbsr} combine a non-blind SR network with cascaded noise and blur kernel estimation blocks to attain a blind architecture. The effectiveness of the methods listed above is limited to the parameterized space of Gaussian kernels and they neglect complex nonlinear motion blurs.  \citet{b31} claim that frequency domain is more conducive than spatial domain for kernel estimation and reconstructs the blur kernel from the LR image’s spectrum by utilizing kernel’s shape structure. Despite the good results for both Gaussian and motion blur kernels, this method also does not take noise model into account, which is imperative for a robust and generalized solution.

\subsection{Cascaded / Iterative Refinement Approach}
In many different vision tasks, researchers have adopted cascaded and iterative modules to improve the performance of their models, such as video segmentation \citep{locglob}, image restoration \citep{rain}, trajectory prediction \citep{hollstm}, abductive reasoning \citep{abdreason}, point cloud completion \citep{casrefine}, pose estimation \citep{caspyramid}. For instance, \citet{abdreason} generate an initial description for each event representation and iteratively refine it using cascaded decoders. \citet{locglob} study cascaded, multi-stage encoding of visual representations for better video segmentation. Likewise, \citet{casrefine} initially perform a coarse reconstruction on the input point cloud and apply iterative refinement on the intermediate output with its secondary module to achieve a high quality dense reconstruction, similar in spirit to our approach.

When it comes to image super-resolution, several non-blind model-based approaches \citep{dpsr, b15} were proposed exploiting the advantages of iterative refinement idea,  besides the aforementioned learning based methods \citep{b13, b14}. \citet{dpsr} decouple the problem into deblurring and SR tasks. The first task is solved in closed form in Fourier domain while the second task is handled by a non-blind SR network. USRNet \citep{b15} also unfolds the objective function which allows to embed the degradation information into the learning model effectively. 
In contrast with DPSR \citep{dpsr}, USRNet  decouples optimization problem into SR reconstruction and denoising tasks. USRNet is both a model-based and end-to-end trainable iterative architecture that accepts blur kernel, noise variance, and scale factor as input parameters in order to provide a general solution to the non-blind SISR problem.

 In this paper, we propose to transform USRNet into a blind SISR network by integrating kernel and noise estimation into the iterative reconstruction process. We claim that combining model-based reconstruction and learning-based correction is essential for robust SR results. These two components of the model can be seen as providing complementary knowledge of the underlying problem: model-based module is responsible for the degradation process and learning-based module learns how to handle the denoising process. This is similar to the  concept of multiple knowledge representation, which is introduced in \citet{mkr} and shown to provide a more robust and regularized feature representation of the underlying data and improve the generalization capability of the model.

  This paper builds upon our previous work to achieve state-of-the-art results in blind SISR.   
  In BISR-Net \citep{b16}, we introduce a kernel initializer model that provides an initial kernel estimate and a kernel estimator model that iteratively corrects the estimated kernel. In \citet{b34}, we decouple the iterative kernel refinement process into model-based reconstruction and learning-based correction sub-modules. In this paper, we extend our previous work by integrating noise estimation into the iterative framework as well. As a result, we claim that the model becomes robust to any type of degradation in the LR image and achieve state-of-the-art performance for isotropic/anisotropic Gaussian and linear/nonlinear motion blurs with different levels of noise.

\begin{figure*}[t]
\centering
\includegraphics[scale = 0.7]{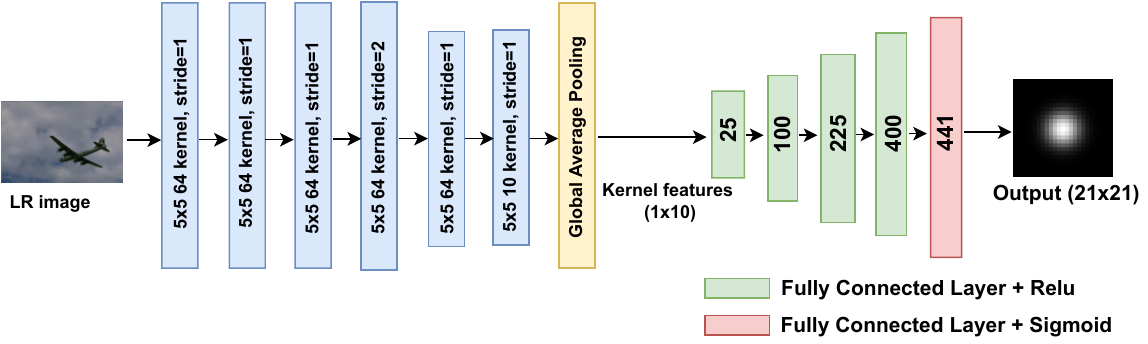}
\caption{Kernel initializer $\mathcal{I}$ takes LR image as input and generates an initial guess $k_0$ for the kernel.}
\label{fig3}
\end{figure*}

\section{Proposed Method}
In this paper, we consider the blind SR problem formulated as in Equation \ref{eq1}. Due to the ill-posed nature of SISR problem, a constrained optimization approach is applied for the solution:
\begin{equation}
\hat{x}, \hat{k} = \arg\min_{x,k} || y-(x\otimes k)\downarrow_s||^2_2+ \lambda \Phi(x)+\Theta(k)
\label{eq2}
\end{equation}
Here, $\Phi(x)$ and $\Theta(k)$ are the regularization functions that represent the prior information about the image and the kernel, respectively; $\lambda$ denotes the Lagrangian multiplier that depends on the noise variance. To solve this problem iteratively, half quadratic splitting (HQS) method can be applied in order to break down the equation  into following steps:
\begin{equation}
\begin{aligned}
k_{i} &= \arg\min_{k} || y-(x_{i-1}\otimes k)\downarrow_s||^2_2+\Theta(k) \\
z_{i} &= \arg\min_{z} || y-(z\otimes k_{i})\downarrow_s||^2_2+\mu ||z-x_{i-1}||^2_2 \\
x_{i} &= \arg\min_{x} \mu ||z_{i}-x||^2_2 + \lambda \Phi(x)
\end{aligned}
\label{eq3}
\end{equation}
As discussed in BISR-Net \citep{b16}, three different modules are implemented to realize an iterative solution for this multi-layer optimization problem. In \citep{b34} and in this paper, instead of using a deep network module to update the kernel estimate, we utilize the HQS technique for the kernel correction step as well:
\begin{equation}
\begin{aligned}
w_{i} &= \arg\min_{w} || y\!-\!(x_{i-1}\otimes w)\downarrow_s||^2_2\!+\!\zeta ||w\!-\!k_{i-1}||^2_2 \\
k_{i} &= \arg\min_{k} \zeta||w_{i}-k||^2_2+\Theta(k) 
\end{aligned}
\label{eq4}
\end{equation}
where \emph{w} and \emph{z} are auxiliary variables that are used to separate the optimization of \emph{x} and \emph{k} into two consecutive steps. We know that the estimation of $z_{i}$ is a convex problem and has a closed form solution in Fourier domain \citep{b15}. This property can be further exploited for the solution of $w_{i}$, since the equations for $z_{i}$ and $w_{i}$ have similar form (note that convolution is a commutative operation). Moreover, it can be noticed that the equations of $k_{i}$ and $x_{i}$ are actually both denoising problems that can be solved using deep denoiser models. Hence the consecutive optimization steps in Equations \ref{eq3} and \ref{eq4} can be solved with the following modules:

\begin{equation}
\begin{aligned}
w_{i} &= \mathcal{D}k(y, x_{i-1}, k_{i-1}) \\
k_{i} &= \mathcal{P}k(w_{i}) \\
z_{i} &= \mathcal{D}(y, x_{i-1}, k_{i})  \\
x_{i} &= \mathcal{P}(z_{i},\beta_{i})
\end{aligned}
\label{eq5}
\end{equation}
where $\mathcal{D}$ and  $\mathcal{D}k$ are fixed modules that implement the closed form solutions of the corresponding convex optimizations, $\mathcal{P}$ and  $\mathcal{P}k$ are trainable deep networks used for denoising and artifact removal. Thus, kernel estimation is divided into two submodules: a non-trainable module $\mathcal{D}k$ to reconstruct the updated kernel and a kernel denoiser module $\mathcal{P}k$ to apply regularization on the reconstructed kernel. Additionally, a kernel initializer network $\mathcal{I}$ is used to provide the initial estimate $k_{0}$ for the first iteration: 
\begin{equation}
k_{0} = \mathcal{I}(y) 
\end{equation}
LR image is bilinear interpolated to produce the initial SR estimate $x_{0}$. The model hyper-parameters  (i.e. $\beta, \lambda, \mu, \zeta$ and $\alpha$ from Equation \ref{eq6_0}) depend on the noise variance $\sigma^2$ and the level of uncertainty in the reconstruction process. $\beta$ determines the denoising strength of module $\mathcal{P}$ and is updated at each iteration as well. In this paper, we investigate both scenarios where noise variance is zero/fixed or unknown. In case the noise variance is unknown, a noise estimator module $\mathcal{F}$ is also used to iteratively estimate the variance and update the related hyper-parameters $\beta, \alpha$:
\begin{equation}
\label{eq7new}
(\sigma_{i}, \beta_{i}, \alpha_{i})  = \mathcal{F}(y, x_{i-1}, k_{i}) 
\end{equation}
Figure \ref{fig2} summarizes these modules and  iterative updates applied on estimated kernel and SR image. 

\begin{figure*}[t]
\centering
\includegraphics[scale = 0.7]{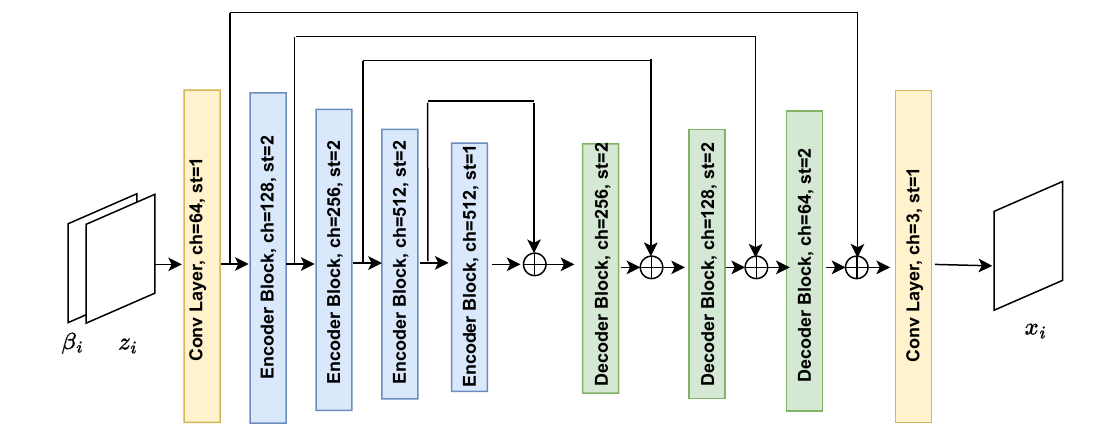}
\caption{ResUNet architecture for SR image denoising module $\mathcal{P}$.}
\label{fig_resunet}
\end{figure*}

\begin{figure*}[t]
\centering 
\includegraphics[scale = 0.7]{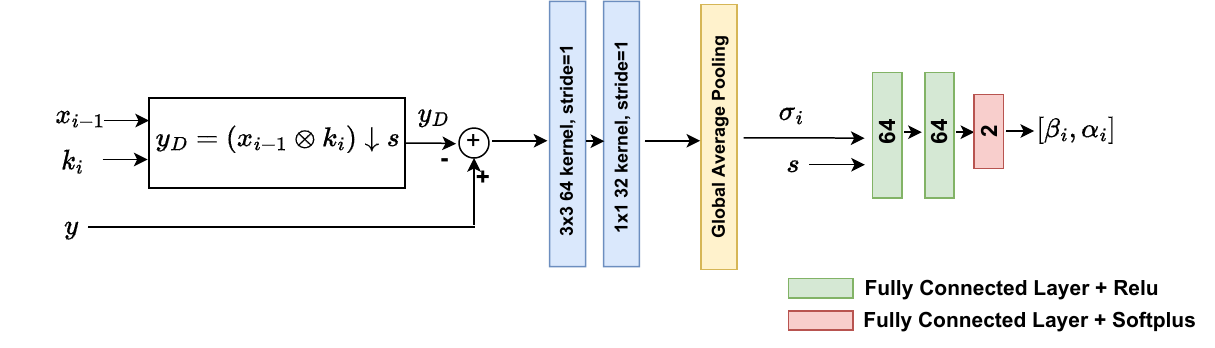}
\caption{Noise and hyper-paramater estimation module $\mathcal{F}$.}
\label{fig_noise_est}
\end{figure*}

\subsection{Kernel Initializer}
Module $\mathcal{I}$ performs an initial estimation of the blur kernel; then the initial kernel is fed to SR reconstruction and kernel reconstruction modules. The architecture of $\mathcal{I}$ is basically an auto-encoder network where a fully convolutional encoder learns the latent feature subspace that uniquely represents different kernels and a fully connected decoder network translates these compressed features into $21\times 21$ kernels. The initializer network is trained in feature subspace via utilizing a separate kernel encoder-decoder model as explained in \citep{b16}. Figure~\ref{fig3} shows the network architecture for the initializer.
 
\subsection{SR Reconstruction  Model}
As depicted in Figure \ref{fig2}, SR Reconstruction model is divided into two sub modules: fixed operator $\mathcal{D}$ and a deep denoiser model $\mathcal{P}$.Module $\mathcal{D}$ performs inverse filtering operation on LR image to reconstruct the SR estimate. Any reconstruction artifacts/noise caused by this inverse filtering are removed by the trainable module $\mathcal{P}$. By separating SR reconstruction into two consecutive steps, it becomes possible to train module $\mathcal{P}$ as a generic denoising network that is almost independent of the underlying blur kernel. In other words, $\mathcal{D}$ is responsible for correct filtering based on the kernel estimate and  $\mathcal{P}$ handles noise removal from the SR estimate of $\mathcal{D}$ .

The solution of $\mathcal{D}$ has a closed form when implemented in Fourier domain, as explained in \citep{b15}: 
\begin{equation}
\begin{aligned}
z_{i} &= \mathcal{F}^{-1}\left(\dfrac{1}{\alpha_i}\Big(d-\overline{\mathcal{F}(k_{i})}\odot_{s}\dfrac{(\mathcal{{F}}(k_{i})d)\!\Downarrow_{s}}{(\overline{\mathcal{F}(k_{i})}\mathcal{F}(k_{i}))\Downarrow_{s}+\alpha_{i}}\Big)\right)
\end{aligned}
\label{eq6_0}
\end{equation}
\begin{equation}
\begin{aligned}
d &= \overline{\mathcal{F}(k_{i})}\mathcal{F}(y\uparrow_{s})+\alpha_{i}\mathcal{F}(x_{i-1})
\end{aligned}
\label{eq7_0}
\end{equation}
Here  $\alpha_i$ is a regularizing constant that is estimated at each iteration by equation \ref{eq7new}, $\mathcal{F(.)}$ denotes FFT, $\odot_{s}$ denotes applying element-wise multiplication to the $s \times s$ distinct blocks of $\overline{\mathcal{F}(k_{i})}$, $\Downarrow_{s}$
denotes averaging the $s \times s$ distinct blocks, $\uparrow_{s}$ denotes  upsampling the spatial size by filling the new entries with zeros.

The denoiser module $\mathcal{P}$ has a CNN architecture based on ResUNet, which is a combination of ResNet \citep{b17} and U-Net \citep{b18} (see \citep{b15} for more details). The module  $\mathcal{D}$ outputs the reconstructed image \emph{z}, which is input to the denoiser module $\mathcal{P}$ for regularization. The denoising strength $\beta$ is concatenated with \emph{z} as an additional constant input channel for $\mathcal{P}$. Then the SR estimate \emph{x} is fed back to $\mathcal{D}$ for iterative update. The architecture for module $\mathcal{P}$ is given in Figure \ref{fig_resunet}. 

Similar to UNet, ResUNet downscales and then upscales the feature maps through encoder and decoder layers, respectively. There is identity skip connection between corresponding layers of the encoder and decoder at four different scales.  In Figure \ref{fig_resunet}, "ch" represents the number of output channels in each layer, and "st" is the stride. Transposed convolution is used at the decoder for upscaling. Each encoder and decoder layer contains two consecutive residual blocks, in addition to the strided and transpose convolution layers, respectively. A residual block is composed of two 3×3 convolution layers with ReLU activation in the middle and an identity skip connection added to its output. The architecture also has two additional convolution layers without any nonlinear activation, one at the input and the other at the output of the network.

\subsection{Kernel Reconstruction  Model}
In Figure \ref{fig2}, kernel reconstruction module also consists of a a fixed $\mathcal{D}k$  and a trainable denoiser/regularizer module $\mathcal{P}k$, just like SR model. Since the convolution of an image with kernel \emph{k} is a commutative operation, the same technique that is used for SR reconstruction can be also applied here. The closed-form solution of $\mathcal{D}k$ is given in Equations \ref{eq6} and \ref{eq7}:
\begin{equation}
\begin{aligned}
w_{i+1} &= \mathcal{F}^{-1}\left(\dfrac{1}{\gamma}\Big(d-\overline{\mathcal{F}(x_{i})}\odot_{s}\dfrac{(\mathcal{{F}}(x_{i})d)\Downarrow_{s}}{(\overline{\mathcal{F}(x_{i})}\mathcal{F}(x_{i}))\!\Downarrow_{s}+\gamma}\Big)\right)
\end{aligned}
\label{eq6}
\end{equation}
\begin{equation}
\begin{aligned}
d &= \overline{\mathcal{F}(x_{i})}\mathcal{F}(y\uparrow_{s})+\gamma\mathcal{F}(k_{i})
\end{aligned}
\label{eq7}
\end{equation}
Here a constant value of $\gamma$ is used for regularization, as opposed to the iteratively updated parameter $\alpha$ in module $\mathcal{D}$. 

The regularizer $\mathcal{P}k$ is also a ResUNet architecture, as in SR reconstruction model, but without the additional input channel for the denoising strength. Module $\mathcal{P}k$  removes noise/artifacts in the kernel estimate $w$ of module $\mathcal{D}k$. Then the updated kernel \emph{k} is fed back to $\mathcal{D}k$ for iterative refinement.

\subsection{Noise Estimator Module}
The noise and hyper-parameter estimator module is given in Figure \ref{fig_noise_est}. The module takes LR image, SR estimate and kernel estimate as input. The SR estimate is filtered with the kernel and downsampled to produce denoised LR image, $y_{D}$:
\begin{equation}
y_{D} = (x_{i-1}\:\otimes\:k_i)\downarrow_{s} 
\end{equation}
The difference between the noisy LR image $y$ and the denoised output $y_{D}$ is input to a 2-layer convolution network 
in order to estimate the noise standard deviation $\sigma$. This standard deviation is used together with the scale factor $s$ to adjust the hyper-parameters $\alpha$ and $\beta$ through a simple 3-layer fully-connected network. 

\section{Experiments}
This section first provides the details on implementing data pre-processing and network training, and then compares the proposed IKR-Net model to other existing blind  SR models.
\subsection{Details of Implementation and Training}
For the training the model, DIV2K \citep{b19} and Flickr2K \citep{b20} datasets consisting of total 3450 HR images are used. LR images are generated by using the degradation process in Equation \ref{eq1} with and without noise and for scale factors $\times4$ and $\times2$. 100K random Gaussian kernels (isotropic and anisotropic kernels with standard deviations in the range [0.2 4.0] for both axes and rotated by a random angle in $[-\pi,\pi]$) and 100K randomly generated motion kernels with different levels of nonlinearity are used for blurring. Also, data augmentation techniques such as random cropping, rotation and flipping are applied to the training images. Two different models are trained for scales $\times4$ and $\times2$. 

Kernel initializer $\mathcal{I}$ is initially trained using L2 loss on 10-dimensional latent feature space generated by a trained kernel encoder (see \citep{b16} for more details) and then fine tuned using L2 loss on the predicted kernel coefficients. SR reconstruction model is first trained with the initial kernel estimate of $\mathcal{I}$ and without using kernel reconstruction model. L1 loss is used for SR model training. Noise variance is assumed to be known at this stage and hyper-parameter estimator in Figure \ref{fig_noise_est} is trained together with module $\mathcal{P}$. After $\mathcal{P}$ is trained for a number of epochs, kernel regularizer $\mathcal{P}k$ is trained using the generated SR images while $\mathcal{P}$ is frozen. $\mathcal{P}k$ is trained with L2 loss on kernel coefficients. Afterwards, the noise estimator $\mathcal{F}$ is trained using L2 loss on the standard deviation of the noise. The maximum standard deviation of the additive Gaussian noise is assumed to be 3\%. Finally, all modules are integrated as in Figure \ref{fig2} and $\mathcal{P}$ and $\mathcal{F}$ are fine-tuned while freezing $\mathcal{I}$ and $\mathcal{P}k$ modules. The number of iterations in SR reconstruction is set as 16.
For training, ADAM optimizer is used with \ensuremath{\beta_1} = 0.9, \ensuremath{\beta_2} = 0.99, batch size of 8, and learning rate of 10e-4. 

\begin{figure}[htbp]
\centering
{\includegraphics[scale=0.2]{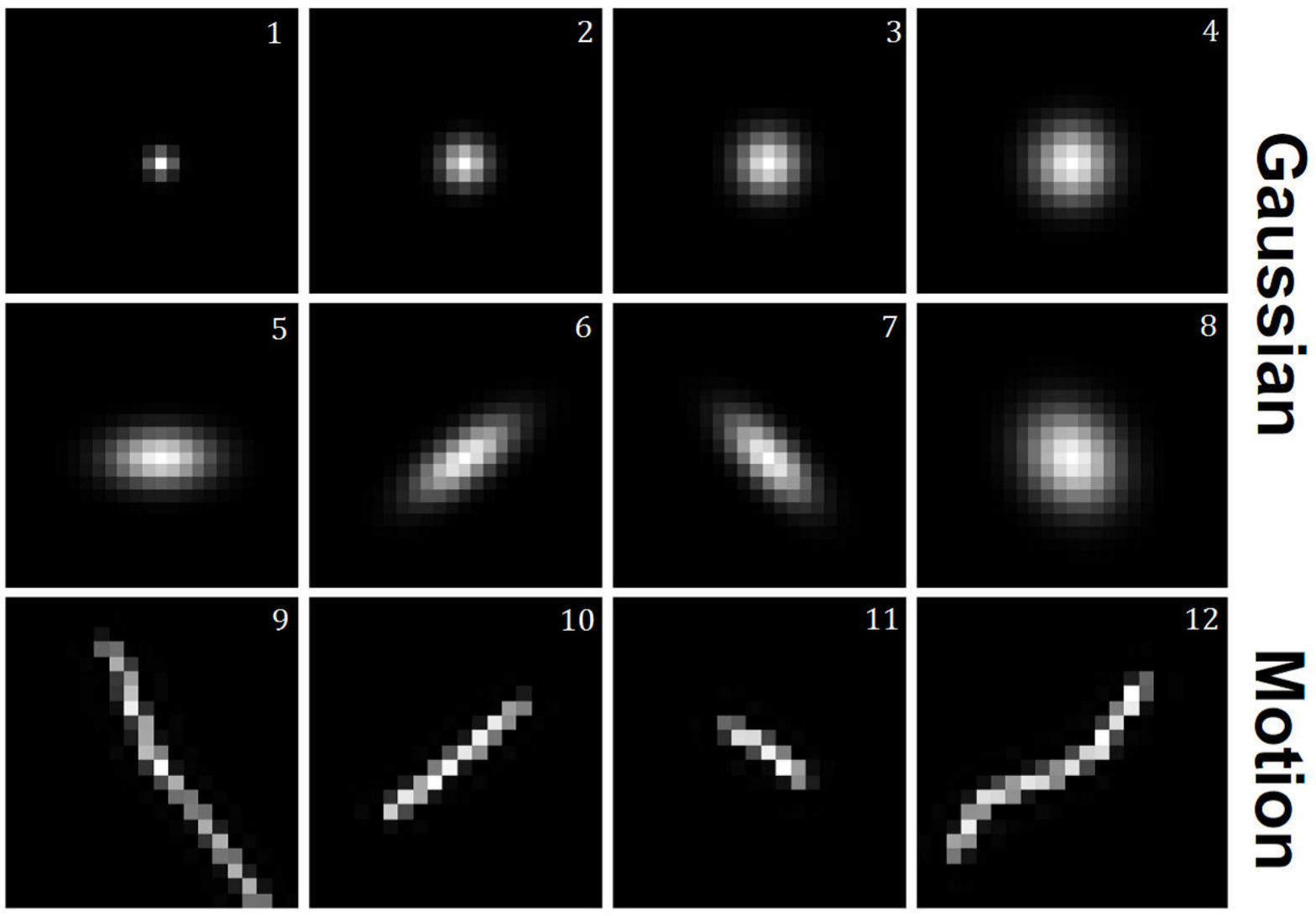}
\caption{Gaussian and motion blur kernels used for testing.}
\label{fig6}}
\end{figure}

\begin{table*}
\caption{PSNR comparison of blind SR models for scale factor $\times 4$ (CBSD68 and Urban100 datasets). For each kernel the best result is highlighted in bold and second-best is highlighted in red.}
\begin{center}
\begin{tabular}{|l| c c c c|c c c c|c c c c |c|c}
\hline
\textbf{CBSD68}&\multicolumn{13}{|c|}{\textbf{Blur Kernels}} \\
 \cline{2-14}
 \textbf{SR Methods}&\textbf{\textit{I}}& \textbf{\textit{II}} & \textbf{\textit{III}}&\textbf{\textit{IV}}&\textbf{\textit{V}} &\textbf{\textit{VI}} & \textbf{\textit{VII}} &\textbf{\textit{VIII}}&\textbf{\textit{IX}}&\textbf{\textit{X}}&\textbf{\textit{XI}}&\textbf{\textit{XII}}& \textbf{Av.}\\
\hline
FKP                & 23.66 & 25.17 & 25.55 & 25.41 & 25.25 & 24.86 & 25.25 & 25.19 & 20.78 & 22.06 & 24.46 & 20.82& 24.04\\
BSRNet              & 24.44 & 24.90 & 24.98 & 24.94 & 24.75 & 24.73 & 24.73 & 24.76 & 21.76 & 23.24 & 24.49 & 21.70& 24.12\\
DASR                & 25.05 & 25.63 & 25.78 & 25.82 & 25.67 & 25.55 & 25.63 & 25.73 & 21.69 & 23.24 & 25.20 & 21.53& 24.71\\
DASR({\it ft})      & 25.09 & 25.53 & 25.53 & 25.44 & 25.43 & 25.38 & 25.45 & 25.30 & 22.45 & 23.86 & 25.22 & 22.40& 24.76\\
IKC({\it ft})     & \textcolor{red}{25.27} & 25.59 & 25.75 & 25.70 & 25.54 & 25.30 & 25.52 & 25.47 & 22.07 & 23.85 & 25.08 & 22.37& 24.79\\
DANv2s1({\it ft}) & \textcolor{red}{25.27} & \textcolor{red}{25.82} & \textcolor{red}{26.00} & \textcolor{red}{26.02} & 25.22 & 25.18 & 25.37 & 25.66 & 22.17 & 23.96 & 25.35 & 22.46& 24.87\\
DANv2s2({\it ft}) & 25.22 & 25.76 & 25.93 & 26.00 & \textcolor{red}{25.86} & \textcolor{red}{25.78} & \textcolor{red}{25.79} & \textcolor{red}{25.98} & 22.31 & 23.97 & \textcolor{red}{25.45} & 22.47& \textcolor{red}{25.04}\\
BISR-Net         & 25.13  & 25.63 & 25.69 & 25.62  & 25.58  & 25.30  & 25.28  & 25.35  & \textcolor{red}{22.50}   & \textcolor{red}{24.52}  & 25.12  & \textcolor{red}{22.88}& 24.88\\
IKR-Net                     & \textbf{25.35}  & \textbf{25.94} & \textbf{26.10} & \textbf{26.12}  & \textbf{25.89}  & \textbf{25.79}  & \textbf{25.84}  & \textbf{26.02}  & \textbf{23.52}  & \textbf{24.71}  & \textbf{25.59}  & \textbf{23.60}& \textbf{25.37}\\
\hline
\hline
\textbf{Urban100}&\multicolumn{13}{|c|}{\textbf{Blur Kernels}} \\
 \cline{2-14}
 \textbf{SR Methods}&\textbf{\textit{I}}& \textbf{\textit{II}} & \textbf{\textit{III}}&\textbf{\textit{IV}}&\textbf{\textit{V}} &\textbf{\textit{VI}} & \textbf{\textit{VII}} &\textbf{\textit{VIII}}&\textbf{\textit{IX}}&\textbf{\textit{X}}&\textbf{\textit{XI}}&\textbf{\textit{XII}}&\textbf{Av.}\\
\hline
BSRNet              & 22.10 & 22.66 & 22.72 & 22.63 & 22.35 & 22.27 & 22.28 & 22.39 & 18.90 & 20.30 & 22.08 & 18.74& 21.62\\
DASR                & 23.06 & 23.60 & 23.62 & 23.55 & 23.38 & 23.23 & 23.29 & 23.35 & 18.90 & 20.23 & 22.71 & 18.58& 22.29\\
DASR({\it ft})      & 22.96 & 23.30 & 23.11 & 22.93 & 22.91 & 22.84 & 22.87 & 22.75 & 19.49 & 21.04 & 22.67 & 19.34& 22.18\\
IKC(ft)           & \textcolor{red}{23.32} & 23.68 & 23.64 & 23.37 & 23.15 & 23.10 & 23.00 & 22.81 & 19.05 & 21.10 & 22.68 & 19.05& 22.33\\
DANv2s1({\it ft}) & 23.20  & 23.77 & \textcolor{red}{24.03} & \textbf{24.08} & 22.95 & 22.57 & 22.84 & 23.64 & 19.37 & 21.14 & 22.82 & 19.48& 22.49\\
DANv2s2({\it ft}) & 23.22  & \textcolor{red}{23.84} & 23.91 & 23.91 & \textbf{23.76} & \textbf{23.63} & \textbf{23.58} & \textbf{23.77} & 19.48 & 21.28 & \textcolor{red}{23.19} & 19.52& \textcolor{red}{22.76}\\
BISR-Net          & 23.17 & 23.57 & 23.35 & 23.02 & 22.97 & 22.70 & 22.73 & 22.75 & \textcolor{red}{19.78} & \textcolor{red}{21.67}  & 22.68  & \textcolor{red}{20.05}& 22.37\\
IKR-Net                     & \textbf{23.93}  & \textbf{24.32} & \textbf{24.22} & \textcolor{red}{23.97}  & \textcolor{red}{23.72}  & \textcolor{red}{23.48}  & \textcolor{red}{23.47}  & \textcolor{red}{23.73}  & \textbf{20.70}  & \textbf{22.13}  & \textbf{23.62}  & \textbf{20.86}& \textbf{23.18}\\
\hline

\end{tabular}
\label{tab_x4}
\end{center}
\end{table*}

\begin{table*}
\caption{PSNR comparison of blind SR models for scale factor $\times 2$ (CBSD68 and Urban100 datasets). For each kernel the best result is highlighted in bold and second-best is highlighted in red.}
\begin{center}
\begin{tabular}{|l| c c c c|c c c c|c c c c |c|c}
\hline
\textbf{CBSD68}&\multicolumn{13}{|c|}{\textbf{Blur Kernels}} \\
 \cline{2-14}
 \textbf{SR Methods}&\textbf{\textit{I}}& \textbf{\textit{II}} & \textbf{\textit{III}}&\textbf{\textit{IV}}&\textbf{\textit{V}} &\textbf{\textit{VI}} & \textbf{\textit{VII}} &\textbf{\textit{VIII}}&\textbf{\textit{IX}}&\textbf{\textit{X}}&\textbf{\textit{XI}}&\textbf{\textit{XII}}& \textbf{Av.}\\
\hline
FKP                    & 24.70 & 28.69 & 28.44 & 26.79 & 26.41 & 25.87 & 26.18 & 25.31 & 15.91 & 14.68 & 16.80 & 15.34& 22.93\\
BSRGAN                  & 27.64 & 27.51 & 27.07 & 26.50 & 26.14 & 25.84 & 25.88 & 25.74 & 21.64 & 22.90 & 25.88 & 21.49& 25.35\\
DASR                    & 29.61 & 29.41 & 28.79 & 27.87 & 27.46 & \textcolor{red}{27.35} & \textcolor{red}{27.55} & 26.76 & 21.80 & 22.86 & 26.06 & 21.71& 26.44\\
DASR({\it ft})          & 29.61 & 29.17 & 28.49 & 27.54 & 27.41 & 27.10 & 27.30 & 26.54 & 22.19 & 24.71 & 27.67 & 22.47& 26.68\\
IKC({\it ft})         & 29.60 & 29.08 & 28.82 & \textcolor{red}{28.18} & \textcolor{red}{27.60} & 26.51 & 25.89 & \textbf{27.34} & \textcolor{red}{22.88} & \textcolor{red}{25.27} & 25.75 & \textcolor{red}{22.68}& 26.63\\
DANv2s1({\it ft})     & 29.92 & 30.03 & 29.08 & 27.51 & 26.04 & 25.56 & 25.67 & 25.98 & 22.27 & 24.44 & 27.84 & 22.46& 26.40\\
DANv2s2({\it ft})     & \textcolor{red}{30.06} & \textcolor{red}{30.28} & \textbf{29.67} & 28.00 & \textbf{27.88} & \textbf{27.76} & \textbf{28.06} & 26.64 & 21.99 & 25.16 & \textbf{28.41} & 22.66 & \textcolor{red}{27.21}\\
BISR-Net              & 29.61  & 29.28 &28.89 & 27.87 & 25.26 & 25.23 & 26.08 & 26.91 & 22.28 & 24.09 & 26.27 & 22.65& 26.20\\
IKR-Net                         & \textbf{30.28} & \textbf{30.36} & \textcolor{red}{29.61} & \textbf{28.25} & 27.58 & 26.66 & 27.02 & \textcolor{red}{27.17} & \textbf{23.29} & \textbf{26.75} & \textcolor{red}{28.31} & \textbf{23.92}& \textbf{27.43}\\
\hline
\hline
\textbf{Urban100}&\multicolumn{13}{|c|}{\textbf{Blur Kernels}} \\
 \cline{2-14}
 \textbf{SR Methods}&\textbf{\textit{I}}& \textbf{\textit{II}} & \textbf{\textit{III}}&\textbf{\textit{IV}}&\textbf{\textit{V}} &\textbf{\textit{VI}} & \textbf{\textit{VII}} &\textbf{\textit{VIII}}&\textbf{\textit{IX}}&\textbf{\textit{X}}&\textbf{\textit{XI}}&\textbf{\textit{XII}}& \textbf{Av.}\\
\hline
BSRGAN           & 25.19 & 24.99 & 24.61 & 24.17 & 23.65 & 23.35 & 23.36 & 23.51 & 18.73 & 19.74 & 22.92 & 18.39& 22.72\\
DASR            & 27.47 & 26.85 & 26.14 & 25.26 & \textcolor{red}{24.81} & \textcolor{red}{24.72} & \textcolor{red}{24.86} & 24.24 & 18.92 & 19.95 & 22.95 & 18.67& 23.74\\
DASR({\it ft})   & 27.36 & 26.51 & 25.83 & 25.01 & 24.78 & 24.49 & 24.66 & 24.05 & 19.43 & 21.54 & 24.70 & 19.34& 23.97\\
IKC({\it ft}) & 27.82 & 26.16 & 25.80 & 24.90 & 24.72 & 23.47 & 24.73 & 23.93 & 19.34 & 20.11 & 22.60 & 19.14& 23.56\\
DANv2s1({\it ft}) & 28.31 & \textbf{28.34} & \textbf{27.49} & \textbf{26.02} & 23.47 & 22.85 & 22.98 & 24.18 & 19.43 & 21.34 & 25.33 & 19.34& 24.09\\
DANv2s2({\it ft}) & \textcolor{red}{28.38} & \textcolor{red}{28.28} & \textcolor{red}{27.45} & \textcolor{red}{25.97} & \textbf{25.59} & \textbf{25.48} & \textbf{25.71} & \textbf{24.56} & 19.30 & \textcolor{red}{22.15} & \textbf{25.93} & \textcolor{red}{19.61}& \textcolor{red}{24.86}\\
BISR-Net     & 27.34 & 26.07 & 25.26 & 24.41 & 22.74 & 23.05 & 23.13 & 23.68 & \textcolor{red}{19.55} & 21.60 & 23.44 & 19.57& 23.32\\
IKR-Net                     & \textbf{28.68} & 27.95 & 26.98 & 25.61 & \textcolor{red}{24.81} & 24.54 & 24.69 & \textcolor{red}{24.39} & \textbf{20.54} & \textbf{24.00} & \textcolor{red}{25.74} & \textbf{20.46}& \textbf{24.87}\\
\hline

\end{tabular}
\label{tab_x2}
\end{center}
\end{table*}

\begin{table*}
\caption{PSNR comparison of blind SR models for 2\% Gaussian noise and scale factors $\times 4$ and $\times 2$ (CBSD68 dataset). For each kernel the best result is highlighted in bold and second-best is highlighted in red.}
\begin{center}
\begin{tabular}{|l| c c c c|c c c c|c c c c |c|c}
\hline
\textbf{$\times 4$}&\multicolumn{13}{|c|}{\textbf{Blur Kernels}} \\
 \cline{2-14}
 \textbf{SR Methods}&\textbf{\textit{I}}& \textbf{\textit{II}} & \textbf{\textit{III}}&\textbf{\textit{IV}}&\textbf{\textit{V}} &\textbf{\textit{VI}} & \textbf{\textit{VII}} &\textbf{\textit{VIII}}&\textbf{\textit{IX}}&\textbf{\textit{X}}&\textbf{\textit{XI}}&\textbf{\textit{XII}}& \textbf{Av.}\\
\hline
BSRNet              & 24.13 & 24.48 & 24.48 & 24.36 & 24.21 & 24.17 & 24.15 & 24.10 & 21.68 & 23.01 & 24.12 & 21.62& 23.71\\
DASR                & 24.86 & \textcolor{red}{25.27} & \textcolor{red}{25.25} & \textcolor{red}{25.05} & \textcolor{red}{24.90} & \textcolor{red}{24.87} & \textcolor{red}{24.96} & \textcolor{red}{24.67} & 21.71 & 23.27 & \textcolor{red}{24.93} & 21.62& 24.28\\
DASR({\it ft})      & \textcolor{red}{24.87} & 25.18 & 25.07 & 24.79 & 24.74 & 24.75 & 24.86 & 24.42 & \textcolor{red}{22.33} & 23.63 & 24.91 & 22.26& \textcolor{red}{24.32}\\
IKC({\it ft})     & 24.55 & 24.61 & 24.32 & 23.83 & 23.64 & 23.57 & 23.75 & 23.17 & 21.28 & 22.87 & 24.16 & 21.46& 23.43\\
DANv2s1({\it ft}) & 24.59 & 24.66 & 24.28 & 23.76 & 23.66 & 23.65 & 23.65 & 23.22 & 21.51 & 23.03 & 24.16 & 21.82& 23.50\\
DANv2s2({\it ft})& 24.59 & 24.73 & 24.39 & 23.90 & 23.78 & 23.81 & 24.00 & 23.38 & 21.93 & 23.14 & 24.46 & 22.03& 23.68\\
BISR-Net         & 24.68 & 24.77 & 24.36 & 23.82 & 23.70 & 23.88 & 23.77 & 23.25 & 21.94 & \textcolor{red}{23.65} & 24.35 & \textcolor{red}{22.44}& 23.72\\
IKR-Net                     & \textbf{25.03} & \textbf{25.47} & \textbf{25.46} & \textbf{25.27} & \textbf{25.01} & \textbf{25.00} & \textbf{25.11} & \textbf{24.92} & \textbf{22.60} & \textbf{24.03} & \textbf{25.10} & \textbf{22.59}& \textbf{24.63}\\
\hline
\hline
\textbf{$\times 2$}&\multicolumn{12}{|c|}{\textbf{Blur Kernels}} \\
 \cline{2-14}
 \textbf{SR Methods}&\textbf{\textit{I}}& \textbf{\textit{II}} & \textbf{\textit{III}}&\textbf{\textit{IV}}&\textbf{\textit{V}} &\textbf{\textit{VI}} & \textbf{\textit{VII}} &\textbf{\textit{VIII}}&\textbf{\textit{IX}}&\textbf{\textit{X}}&\textbf{\textit{XI}}&\textbf{\textit{XII}}& \textbf{Av.}\\
\hline
BSRGAN              & 27.32 & 26.94 & 26.29 & 25.64 & 25.50 & 25.18 & 25.25 & 24.92 & 21.68 & 23.10 & 25.67 & 21.63& 24.93\\
DASR                 & 28.92 & \textcolor{red}{27.99} & \textcolor{red}{26.95} & \textcolor{red}{26.08} & \textcolor{red}{26.09} & \textbf{25.83} & \textcolor{red}{26.00} & \textcolor{red}{25.34} & 21.71 & 23.06 & 26.31 & 21.66& 25.50\\
DASR({\it ft})       & \textcolor{red}{28.96} & 27.95 & 26.89 & 26.00 & 26.08 & 25.79 & 25.97 & 25.24 & \textcolor{red}{22.11} & \textcolor{red}{24.44} & \textcolor{red}{26.97} & \textcolor{red}{22.42}& \textcolor{red}{25.74}\\
IKC({\it ft})     & 27.63 & 25.94 & 24.86 & 24.10 & 24.09 & 23.84 & 23.99 & 23.45 & 21.60 & 23.01 & 24.95 & 21.69& 24.10\\
DANv2s1({\it ft}) & 27.77 & 26.04 & 24.97 & 24.18 & 24.16 & 23.91 & 24.04 & 23.51 & 21.64 & 23.08 & 25.12 & 21.76& 24.19\\
DANv2s2({\it ft}) & 27.98 & 26.19 & 24.97 & 24.13 & 24.12 & 23.88 & 24.01 & 23.46 & 21.59 & 23.02 & 25.10 & 21.66& 24.18\\
BISR-Net         & 28.16 & 26.41 & 24.98 & 23.95 & 24.02 & 23.90 & 23.75 & 23.17 & 21.45 & 23.38 & 25.10 & 21.80& 24.17\\
IKR-Net                     & \textbf{29.39} & \textbf{28.49} & \textbf{27.34} & \textbf{26.49} & \textbf{26.28} & \textcolor{red}{25.82} & \textbf{26.04} & \textbf{25.75} & \textbf{23.11} & \textbf{25.63} & \textbf{27.40} & \textbf{23.52}& \textbf{26.27}\\
\hline

\end{tabular}
\label{tab_noise}
\end{center}
\end{table*}

\subsection{Evaluation and Comparison}\label{AA}
The proposed model is evaluated on CBSD68 \citep{b21, b22} and Urban100 \citep{b23} datasets, which contain 68 and 100 HR images, respectively. These images are blurred using a set of 12 kernels consisting of both Gaussian and motion blurs. In Figure \ref{fig6}, the first 4 kernels are isotropic, the next 4 are anisotropic Gaussian and the last four are selected motion kernels. The results are compared with blind SISR models of DASR \citep{b30}, BRSGAN/BSRNet \citep{b26}, FKP \citep{b35}, IKC \citep{b13}, DAN \citep{b14}, and BISR-Net \citep{b16}. For fair comparison, the models of DASR, IKC and DAN are fine-tuned with the same set of motion kernels used in IKR-Net and BISR-Net training. BRSGAN/BSRNet models are not fine-tuned, since they claim to be robust against different real-life degradations (note that, while BSRNet model is tested for scale $\times4$, BSRGAN is tested for scale $\times2$ since BSRNet model is not available for this scale). The training of FKP  is unsupervised and thus FKP claims to be applicable for arbitrary kernel assumptions. The second version of DAN model is fine-tuned and tested under two settings: setting-1 for isotropic Gaussians (DANv2s1), and setting-2 for isotropic/anisotropic Gaussians (DANv2s2). DASR model is tested with and without fine-tuning, in order to analyze the performance improvement for motion kernels after fine-tuning. In the following "model({\it ft})" stands for a model fine-tuned with 100K motion kernels.

PSNR results for all tested 12 kernels are presented in Table \ref{tab_x4} for scale $\times4$ and Table \ref{tab_x2} for scale $\times2$, when there is no input noise. IKR-Net outperforms all the tested methods in terms of average PSNR. For scale $\times4$ IKR-Net is the best or second best method in all the tested kernels in both CBSD68 and Urban100 datasets. The PSNR gains are even more impressive for motion kernels.  For scale $\times2$, DANv2s2{\it(ft)} outperforms IKR-Net for anisotropic Gaussian kernels. However  IKR-Net is still the best algorithm in terms of average PNSR. 

Table \ref{tab_noise} presents the test results for noise standard deviation of 2\% in scales $\times4$ and $\times2$. IKR-Net is by far the superior model for noisy images, providing the highest PSNR values in all tested kernels (except for kernel-VI in Urban dataset). DASR stands out as the second best algorithm; the comparison between DASR and DASR{\it(ft)} shows that fine-tuning improves DASR's performance for motion kernels with slight PSNR drop for Gaussian kernels. However fine-tuned DASR still cannot achieve the performance of IKR-Net, especially for motion kernels.

When results in all three tables are analyzed, it is not hard to see that IKR-Net provides the most generalized solution to blind SR problem. IKR-Net achieves consistent and high performance in all tested scales for all isotropic/anisotropic Gaussian and linear/nonlinear motion kernels and at different noise levels. DAN provides competitive performance for isotropic/anisotropic Gaussian when there is no noise; however DAN's performance degrades substantially when there is noise in the tested images. IKC and BISR-Net are also not robust against noise. FKP fails for motion kernels and exhibit very poor results for scale $\times2$. FKP also cannot handle noise in input images and is therefore not included in Table \ref{tab_noise}.  BSRNet/BSRGAN and DASR yield comparatively better results in the presence of noise; yet they also fail to compete with IKR-Net.      

Table \ref{tab_DAN_Setting} compares IKR-Net against existing methods under the test settings provided for DAN in \citep{b14}. In setting-1, 8 isotropic Gaussian kernels are used, with kernel widths uniformly chosen from the range [1.8, 3.2]. No noise is added to the LR images or the kernels. 
In setting-2, anisotropic Gaussian kernels are utilized: the lengths of both axes are uniformly distributed in (0.6, 5), rotated by a random
angle uniformly distributed in $[-\pi, \pi]$. To deviate from a regular Gaussian, uniform multiplicative noise (up to 25\% of each pixel value of the kernel) is applied  and the kernel is normalized to sum to one. LR image is again noise-free. The table reports average PSNR values for the Y-channel (i.e. PSNR-Y) in addition to the PSNR for the whole color image. The results in the table confirm our previous findings. IKR-Net and DAN  have similar PNSR performance for isotropic/anisotropic Gaussians when there is no noise in LR image. IKR-Net outperforms DAN for the isotropic case and DAN  is slightly better for anisotropic Gaussians kernels that are distorted by multiplicative noise. Other methods exhibit substantially lower performance in both settings.

In Figures \ref{fig7}, \ref{fig8}, \ref{fig9}, visual comparisons of SR results for scale factor $\times4$  also confirm the superiority of our method. In Figure \ref{fig7} for noise-free LR images, IKR-Net achieves the best reconstruction performance for fine high-frequency details (e.g., the texture in the man's shirt and the columns of the building). DANv2s2{\it(ft)} is second best and provides slightly more blurry results than IKR-Net. Other methods exhibit blurry outcomes and/or motion artifacts.  In Figure \ref{fig8} for noisy LR images, IKR-Net shows robust visual performance, strong denoising capability with fine reconstruction of image details (e.g., the windows of the building exterior).  DASR{\it(ft)} and BSRNet also filter out the noise successfully, but BSRNet cannot remove motion blur and  DASR{\it(ft)} outputs more blurry results with some high-frequency artifacts. Also when the reconstructed kernels in Figures \ref{fig7} and \ref{fig8} are compared, it is easy to see that IKR-Net provides much more accurate kernel estimates than BISR-Net for both noise-free and noisy LR images.          

In Figure \ref{fig9}, SR results of the best performing four methods are compared for real blurry images. The images are scaled by $\times4$. The top two figures highlight the superior motion deblurring capability of IKR-Net. Despite being fine-tuned with motion kernels, DASR{\it(ft)} and DANv2s2{\it(ft)} cannot remove motion blur effectively. In the bottom figure, IKR-Net provides the sharpest and most noise/artifact-free result for the text image. Other methods produce slightly more blurry outcomes with shadow artifacts in some of the text characters. Therefore IKR-Net is capable of not only removing motion blur but also providing the sharpest and fine-detailed results in real-life test cases.  

In order to highlight the limitations of the proposed SR model, Figure \ref{fig_fail} provides a visual example in which IKR-Net fails to synthesize correctly the high-frequency details of the HR image. Aliasing artifact are visible on the ground and the texture on the wall is not correctly reconstructed. As future work, we plan to apply generative learning techniques, which are known to provide better representation of such challenging high-frequency patterns. 



\begin{table}
\caption{PSNR comparison of blind SR models for scale factor $\times 4$ using test settings in \citep{b14} (CBSD68 dataset).}
\begin{center}
\begin{tabular}{|l| c |c ||c | c|}
\hline
\multirow{2}{*}{\textbf{SR Methods}} & \multicolumn{2}{|c||}{\textbf{setting-1}} & \multicolumn{2}{|c|}{\textbf{setting-2}}\\
\cline{2-5}
 & \textbf{PSNR} & \textbf{PSNR-Y} & \textbf{PSNR} &\textbf{PSNR-Y}\\
\hline
BSRNet             & 24.71 & 26.32 & 24.18 & 25.74 \\
                            \cline{1-5}
DASR({\it ft})     & 25.20 & 26.56 & 24.66  & 26.02\\
\cline{1-5}
IKC({\it ft})      & 25.40 & 26.75 & 24.75 & 26.10\\
                            \cline{1-5}
DANv2s2({\it ft})  & \textcolor{red}{25.88} & \textcolor{red}{27.23} & \textbf{25.32} & \textbf{26.68} \\ 
                            \cline{1-5}
IKR-Net            & \textbf{26.00} & \textbf{27.35} & \textcolor{red}{25.23} & \textcolor{red}{26.58} \\
\hline
\end{tabular}
\label{tab_DAN_Setting}
\end{center}
\end{table}

\subsection{Ablation Study}\label{HH}
This section analyzes the effects of several critical modules and parameters on the performance of IKR-Net. In particular we investigate kernel reconstruction accuracy, noise variance estimation performance, number of model iterations, the use of iterative kernel update and their affect on the SR image reconstruction performance.   

\begin{figure*}[h]

\centerline{
\includegraphics[scale=0.3]{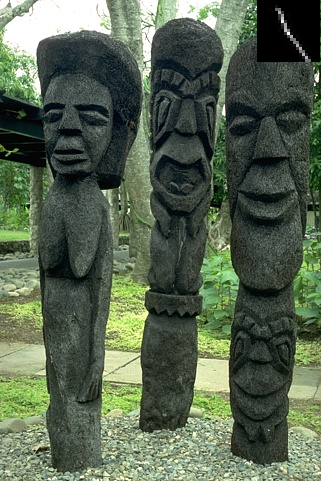}
\includegraphics[scale=0.72]{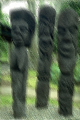}
\includegraphics[scale=0.3]{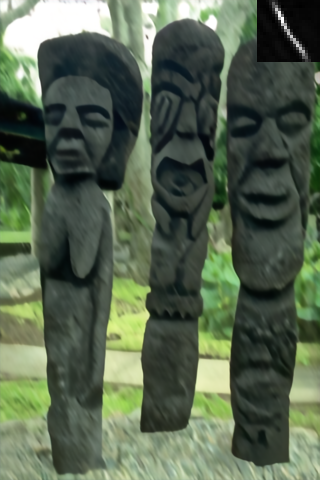}
\includegraphics[scale=0.3]{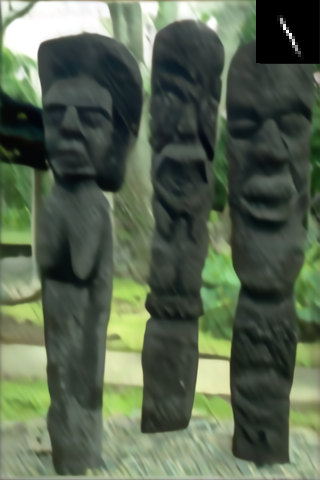}
\includegraphics[scale=0.18]{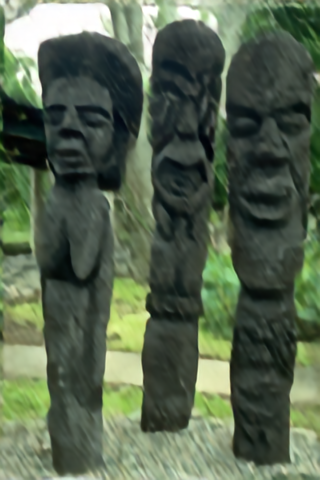}
\includegraphics[scale=0.18]{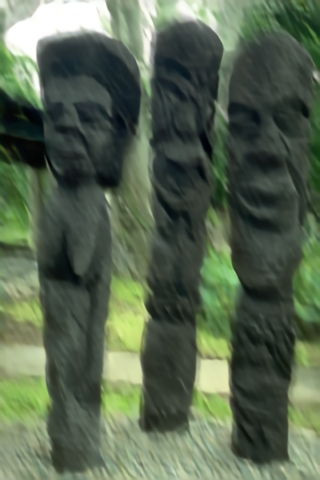}
\includegraphics[scale=0.18]{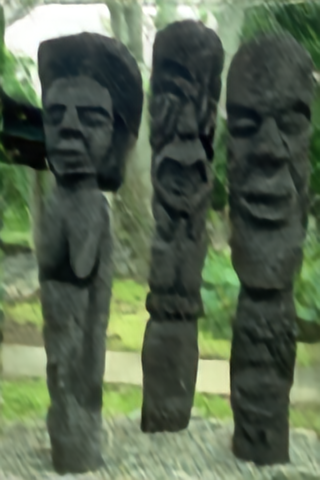}
\includegraphics[scale=0.18]{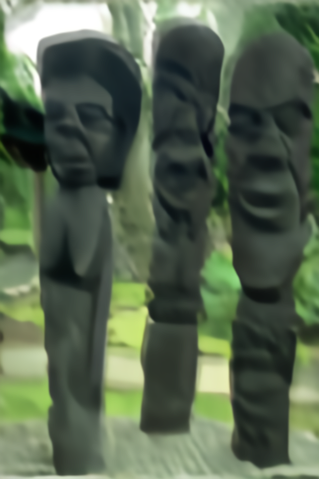}}

\centerline{
\includegraphics[scale=0.3]{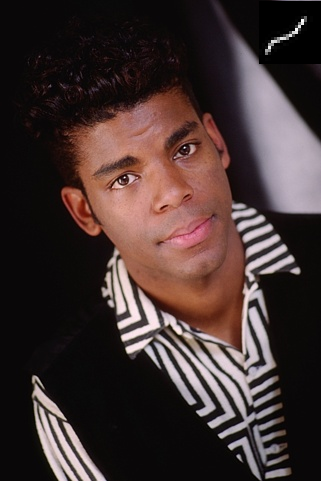}
\includegraphics[scale=0.72]{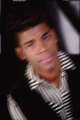}
\includegraphics[scale=0.3]{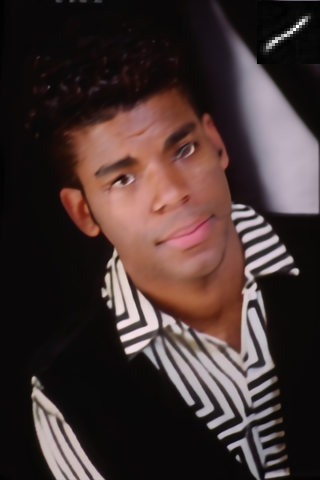}
\includegraphics[scale=0.3]{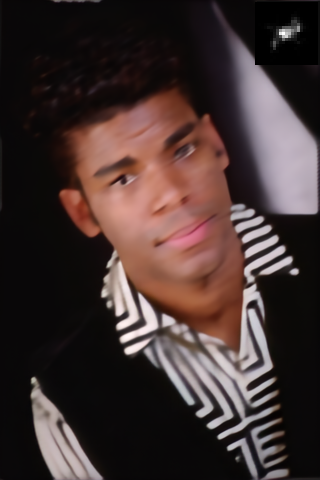}
\includegraphics[scale=0.18]{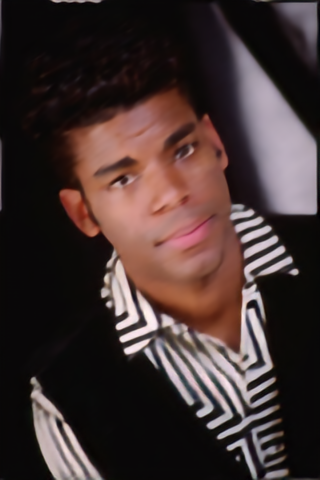}
\includegraphics[scale=0.18]{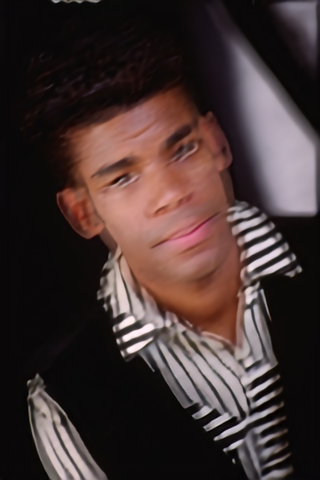}
\includegraphics[scale=0.18]{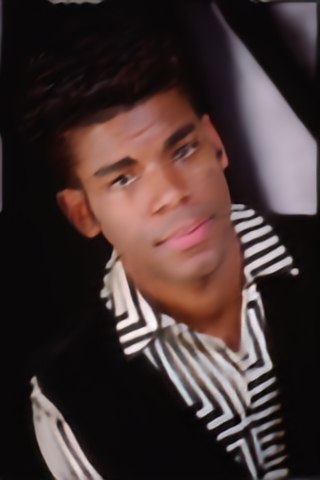}
\includegraphics[scale=0.18]{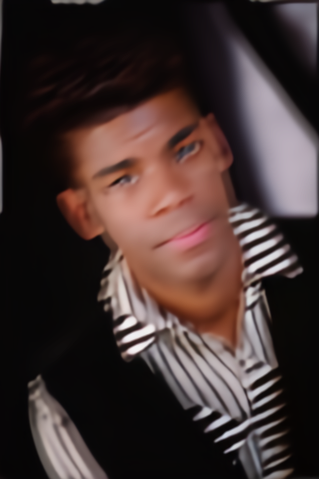}}

\hspace{0.20 cm}\centerline{
\begin{subfigure}{0.125\textwidth}\includegraphics[scale=0.3]{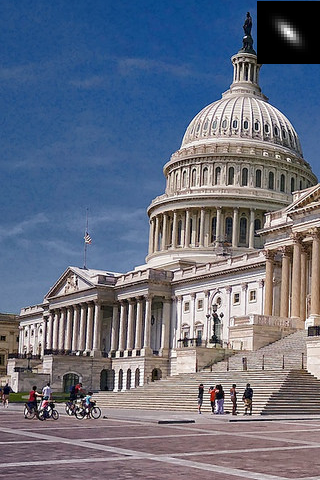}\caption{HR image}\end{subfigure}\hspace{-0.26 cm}
\begin{subfigure}{0.125\textwidth}\includegraphics[scale=0.72]{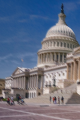}\caption{scaled LR}\end{subfigure}\hspace{-0.26 cm}
\begin{subfigure}{0.125\textwidth}\includegraphics[scale=0.3]{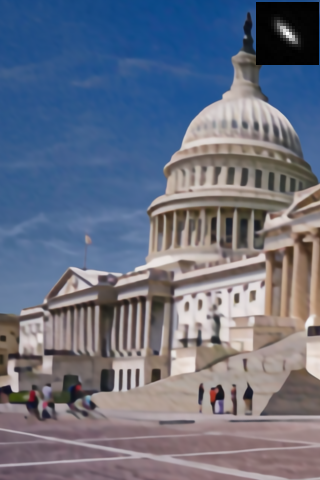}\caption{IKR-Net}\end{subfigure}\hspace{-0.26 cm}
\begin{subfigure}{0.125\textwidth}\includegraphics[scale=0.3]{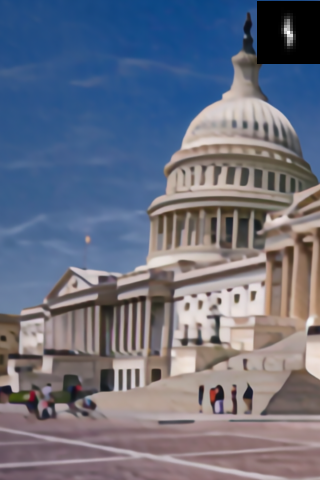}\caption{BISR-Net}\end{subfigure}\hspace{-0.26 cm}
\begin{subfigure}{0.125\textwidth}\includegraphics[scale=0.18]{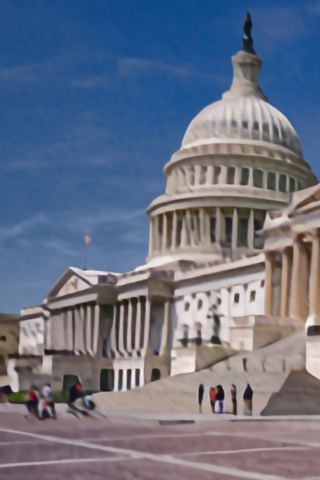}\caption{DANv2s2({\it ft})}\end{subfigure}\hspace{-0.26 cm}
\begin{subfigure}{0.125\textwidth}\includegraphics[scale=0.18]{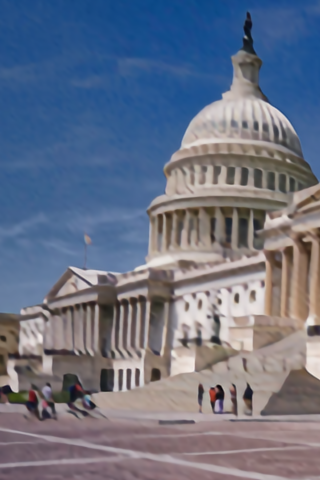}\caption{IKC({\it ft})}\end{subfigure}\hspace{-0.26 cm}
\begin{subfigure}{0.125\textwidth}\includegraphics[scale=0.18]{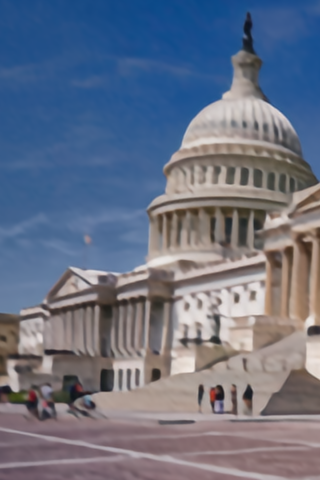}\caption{DASR({\it ft})}\end{subfigure}\hspace{-0.26 cm}
\begin{subfigure}{0.125\textwidth}\includegraphics[scale=0.18]{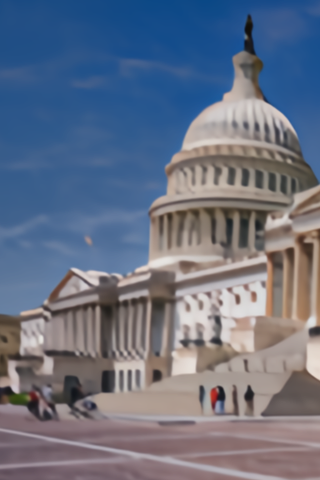}\caption{BSRNet}\end{subfigure}

}

\caption{Visual results of different blind SR methods for noise-free LR images with scale factor $\times 4$. The ground truth blur kernel is shown on the
upper-right corner of the HR image.}
\label{fig7}
\end{figure*}

\begin{figure*}[h]

\centerline{
\includegraphics[scale=0.20]{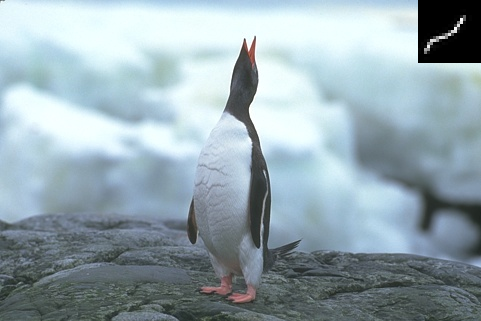}
\includegraphics[scale=0.48]{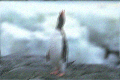}
\includegraphics[scale=0.20]{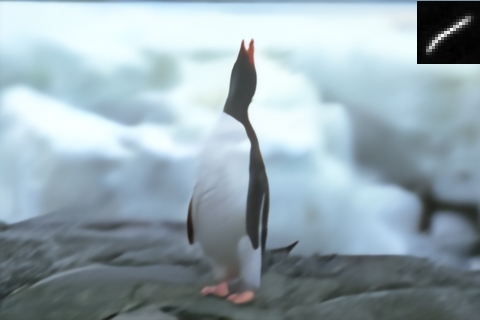}
\includegraphics[scale=0.20]{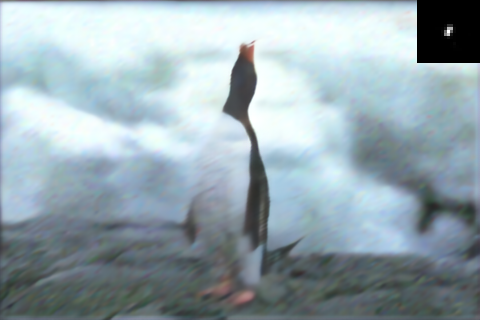}
\includegraphics[scale=0.12]{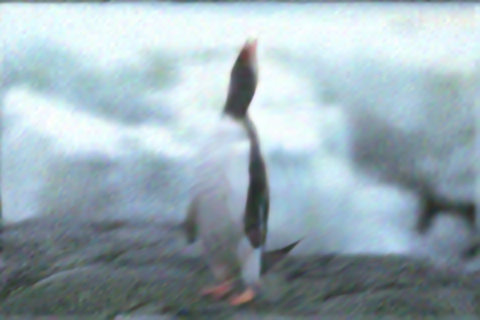}
\includegraphics[scale=0.12]{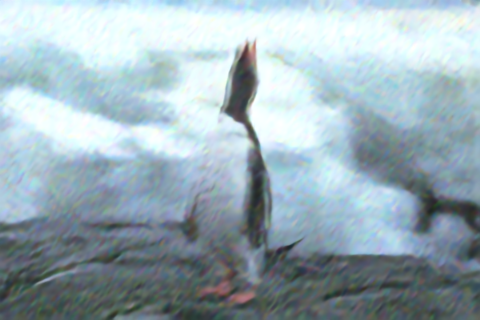}
\includegraphics[scale=0.12]{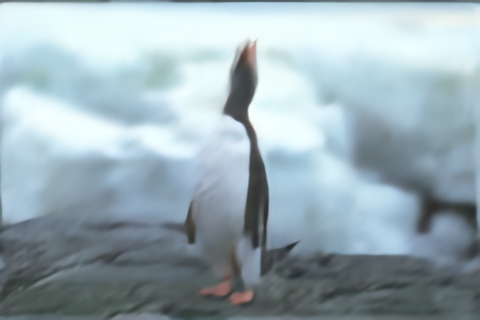}
\includegraphics[scale=0.12]{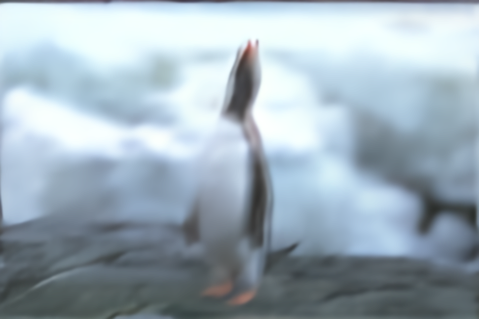}}

\centerline{
\includegraphics[scale=0.3]{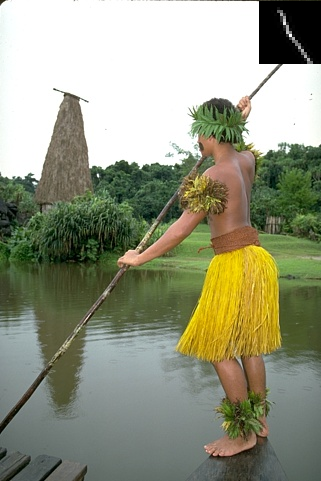}
\includegraphics[scale=0.72]{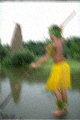}
\includegraphics[scale=0.3]{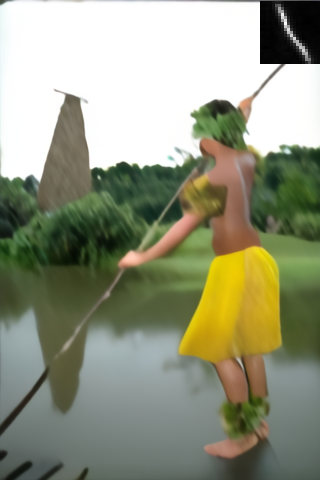}
\includegraphics[scale=0.3]{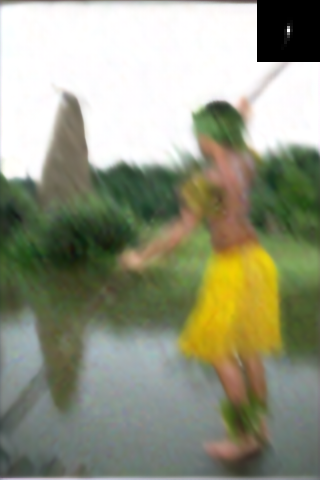}
\includegraphics[scale=0.18]{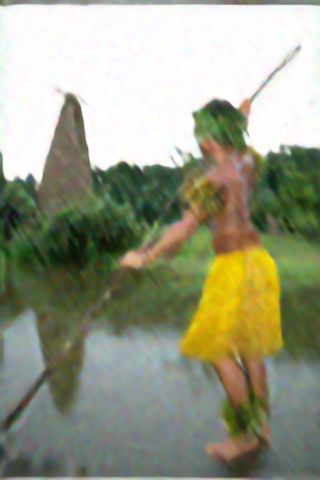}
\includegraphics[scale=0.18]{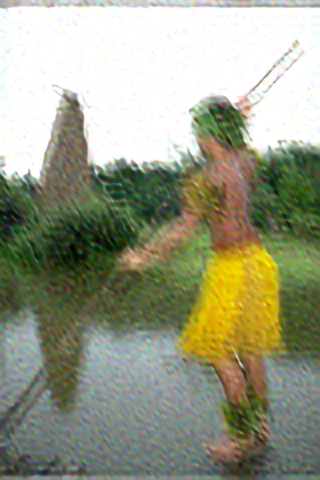}
\includegraphics[scale=0.18]{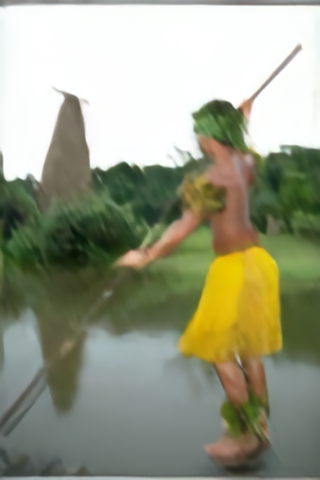}
\includegraphics[scale=0.18]{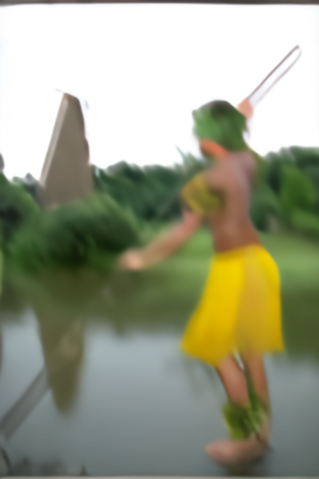}}

\hspace{0.20 cm}\centerline{
\begin{subfigure}{0.125\textwidth}\includegraphics[scale=0.3]{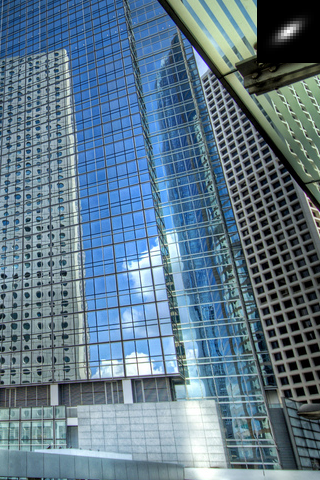}\caption{HR image}\end{subfigure}\hspace{-0.26 cm}
\begin{subfigure}{0.125\textwidth}\includegraphics[scale=0.72]{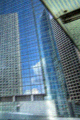}\caption{scaled LR}\end{subfigure}\hspace{-0.26 cm}
\begin{subfigure}{0.125\textwidth}\includegraphics[scale=0.3]{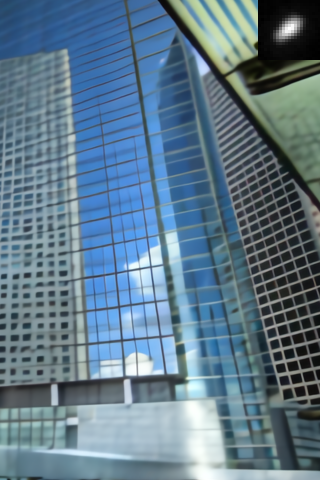}\caption{IKR-Net}\end{subfigure}\hspace{-0.26 cm}
\begin{subfigure}{0.125\textwidth}\includegraphics[scale=0.3]{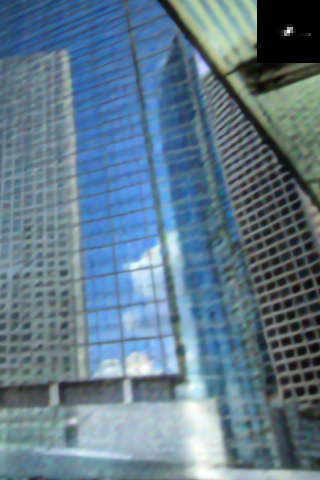}\caption{BISR-Net}\end{subfigure}\hspace{-0.26 cm}
\begin{subfigure}{0.125\textwidth}\includegraphics[scale=0.18]{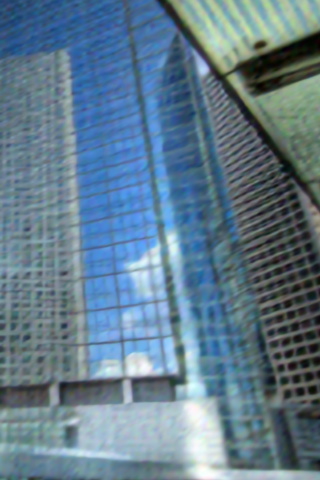}\caption{DANv2s2({\it ft})}\end{subfigure}\hspace{-0.26 cm}
\begin{subfigure}{0.125\textwidth}\includegraphics[scale=0.18]{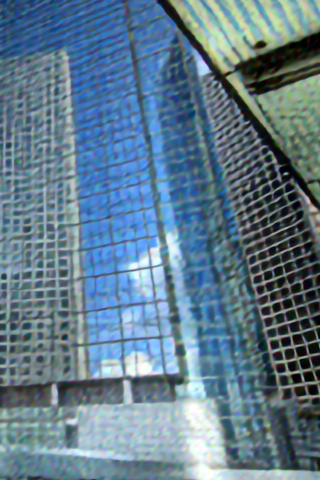}\caption{IKC({\it ft})}\end{subfigure}\hspace{-0.26 cm}
\begin{subfigure}{0.125\textwidth}\includegraphics[scale=0.18]{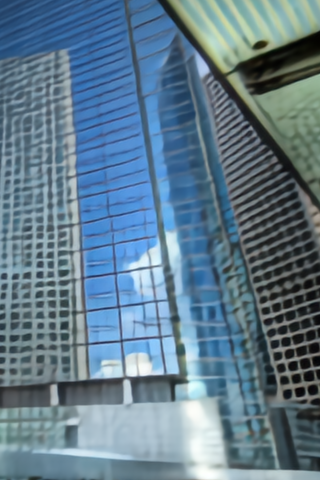}\caption{DASR({\it ft})}\end{subfigure}\hspace{-0.26 cm}
\begin{subfigure}{0.125\textwidth}\includegraphics[scale=0.18]{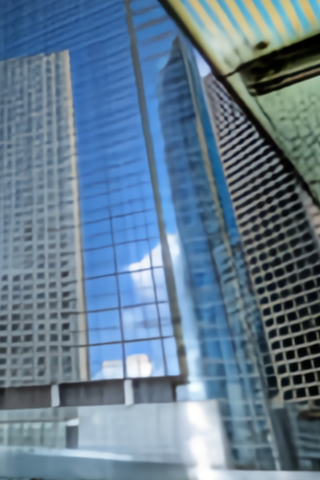}\caption{BSRNet}\end{subfigure}}

\caption{Visual results of different blind SR methods for LR images with 2\% noise and scale factor $\times 4$.}
\label{fig8}
\end{figure*}

\begin{figure*}[h]

\centerline{
\includegraphics[scale=0.33]{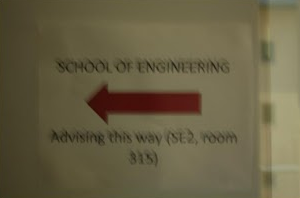}
\includegraphics[scale=0.165]{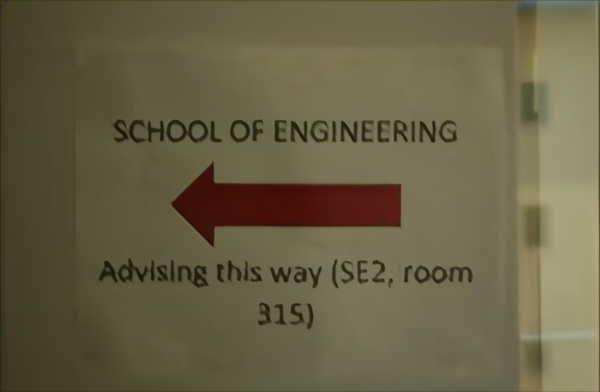}
\includegraphics[scale=0.165]{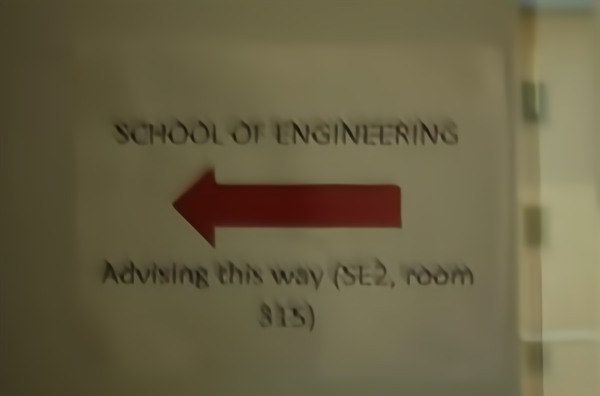}
\includegraphics[scale=0.165]{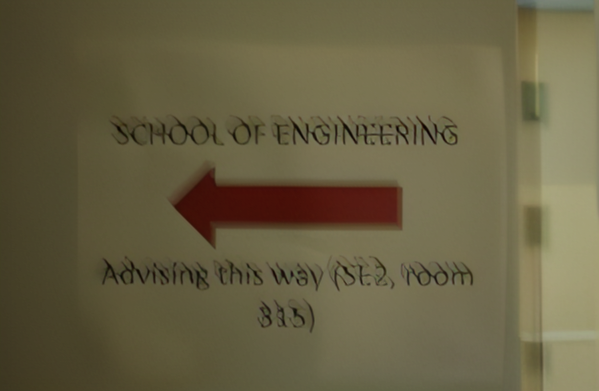}
\includegraphics[scale=0.165]{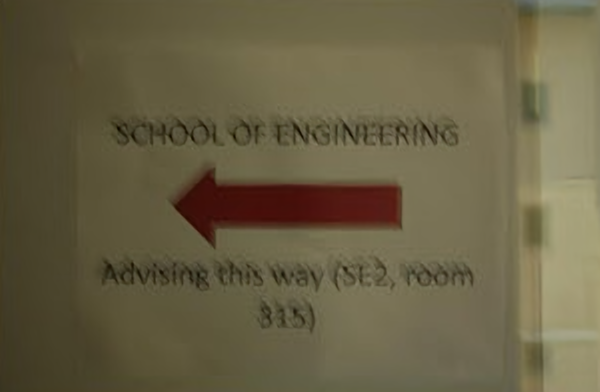}}

\centerline{
\includegraphics[scale=0.52]{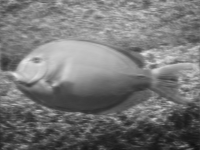}
\includegraphics[scale=0.125]{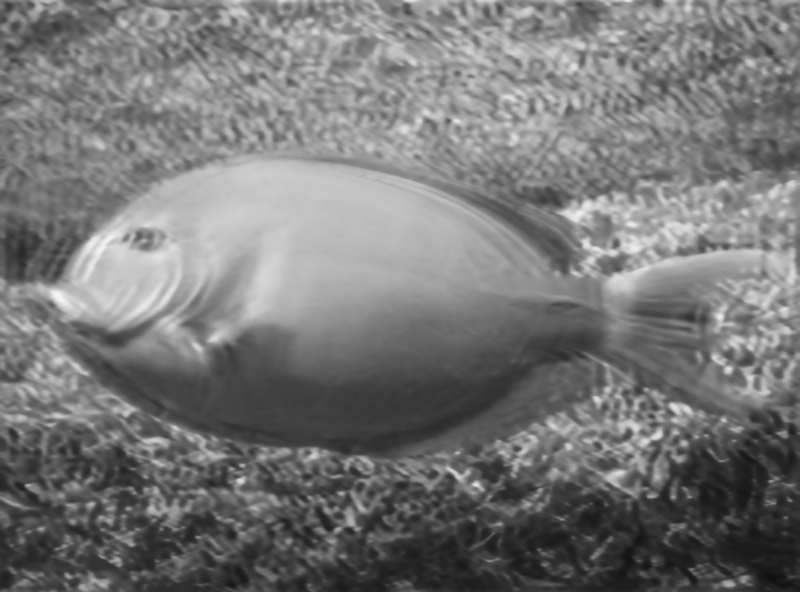}
\includegraphics[scale=0.125]{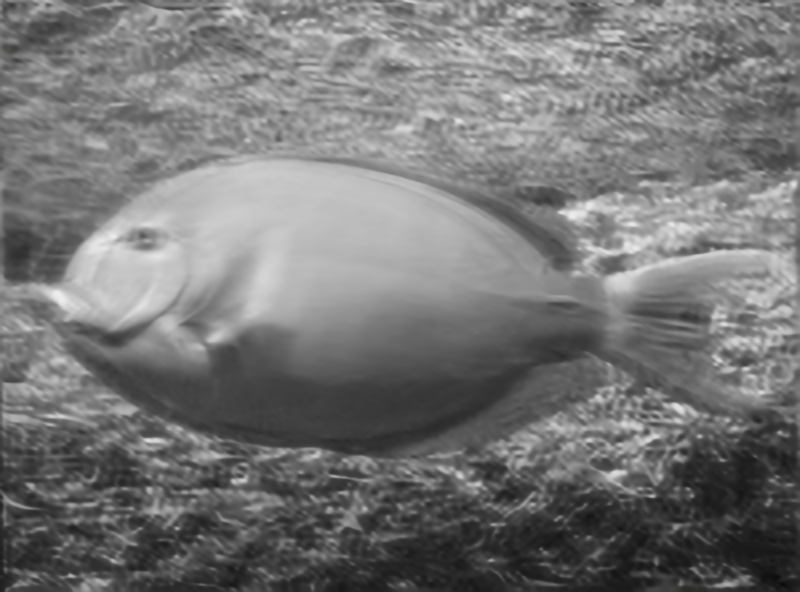}
\includegraphics[scale=0.125]{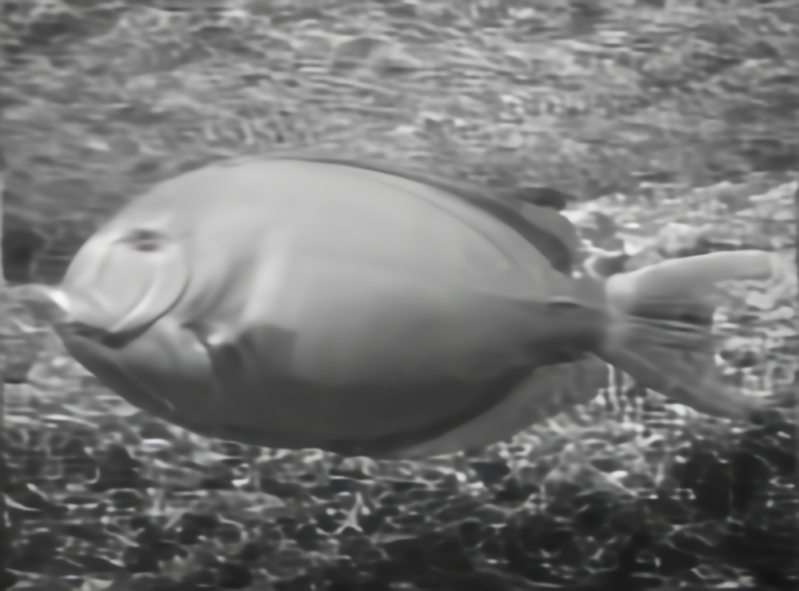}
\includegraphics[scale=0.125]{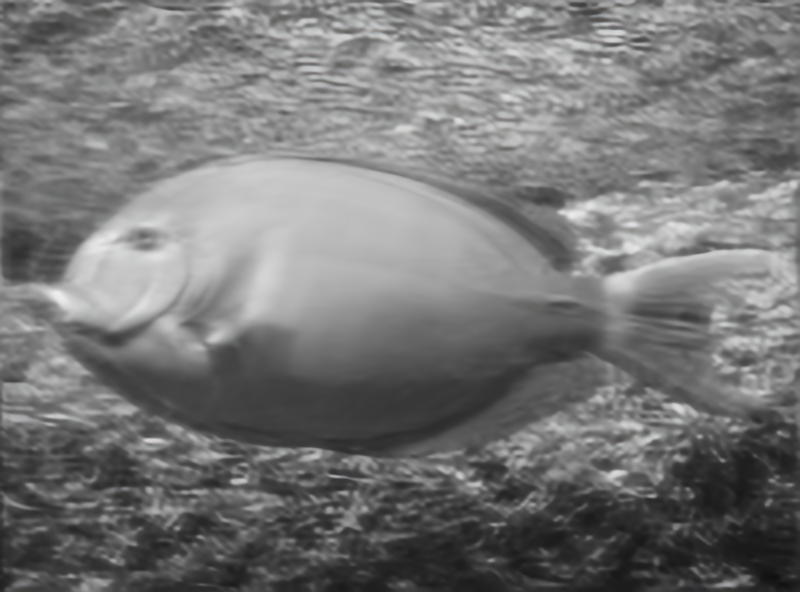}

}

\hspace{0.01 cm}\centerline{
\begin{subfigure}{0.2\textwidth}\includegraphics[scale=0.92]{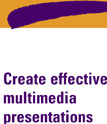}\caption{scaled original}\end{subfigure}\hspace{-0.16 cm}
\begin{subfigure}{0.2\textwidth}\includegraphics[scale=0.23]{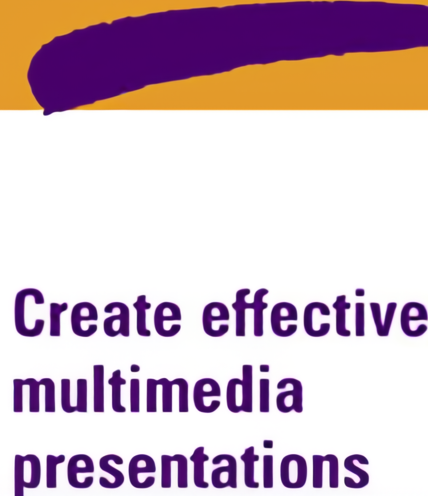}\caption{IKR-Net}\end{subfigure}\hspace{-0.16 cm}
\begin{subfigure}{0.2\textwidth}\includegraphics[scale=0.23]{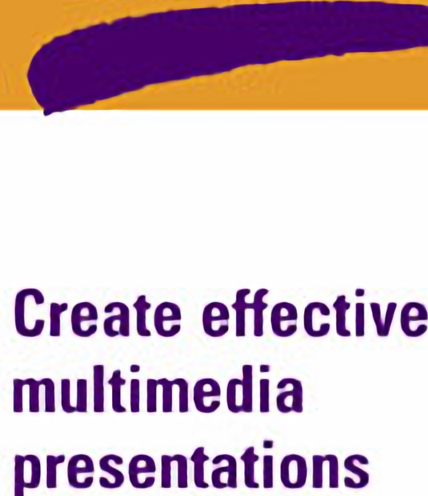}\caption{DANv2s2({\it ft})}\end{subfigure}\hspace{-0.16 cm}
\begin{subfigure}{0.2\textwidth}\includegraphics[scale=0.23]{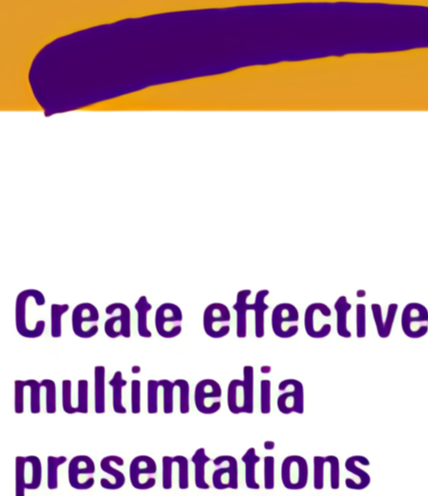}\caption{BSRNet}\end{subfigure}\hspace{-0.16 cm}
\begin{subfigure}{0.2\textwidth}\includegraphics[scale=0.23]{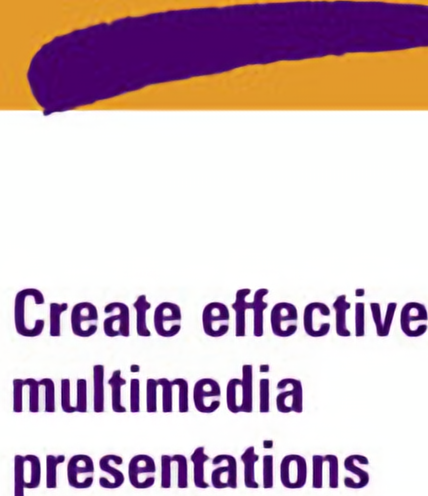}\caption{DASR({\it ft})}\end{subfigure}\hspace{-0.16 cm}

}
\caption{Visual comparisons of different blind SR methods for real test images (scale factor $\times 4$).}
\label{fig9}
\end{figure*}

\begin{figure}[htbp]
\centerline{
\begin{subfigure}{0.35\textwidth}\includegraphics[scale=0.18]{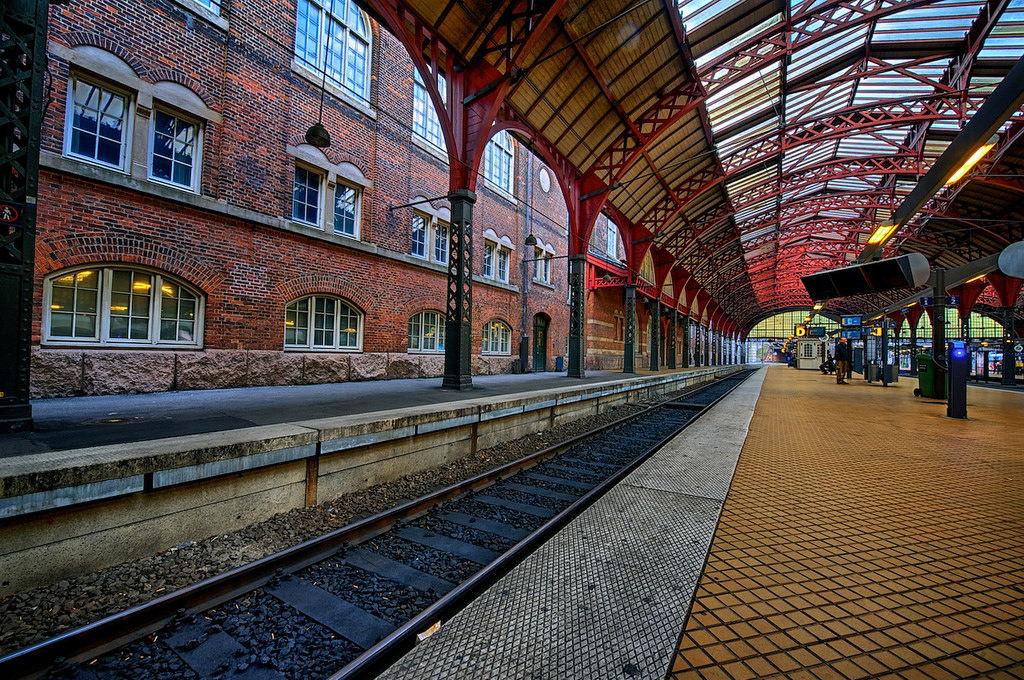}\caption{HR image}\end{subfigure}}
\centerline{\begin{subfigure}{0.35\textwidth}\includegraphics[scale=0.18]{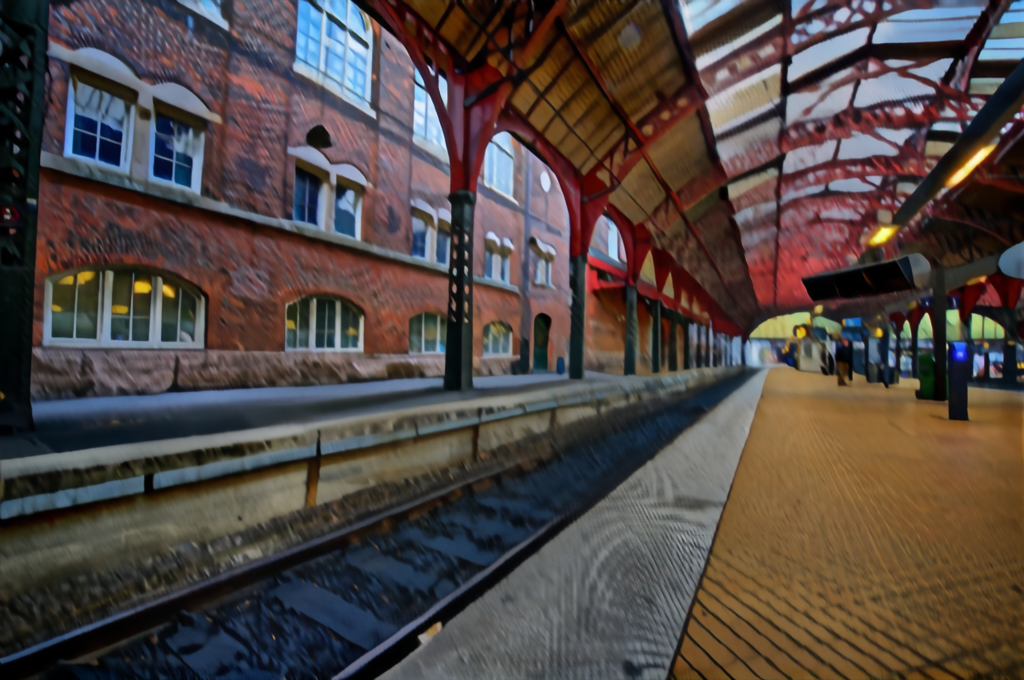}\caption{IKR-Net}\end{subfigure}}
\caption{A test example that exhibits aliasing artifacts and missing high-frequency texture in SR image reconstruction of IKR-Net (scale factor $\times 4$).}
\label{fig_fail}
\end{figure}

Table \ref{tab_ker} gives the average kernel reconstruction error for 12 tested kernels in CBSD68 dataset for scale factor $\times 4$. The models IKR-Net, BISR-Net and FKP are compared in this table, since these three methods provide explicit kernel reconstruction. As the Table implies, and as can be seen from the visual examples in Figure \ref{fig7}, IKR-net provides significantly better kernel estimates, especially for motion kernels, which lead to better SR image reconstruction.

\begin{table*}
\caption{Mean squared error (MSE $\times 10^{-5}$) comparison of kernel estimation for scale factor $\times 4$ (CBSD68 dataset).}
\begin{center}
\begin{tabular}{|l| c c c c|c c c c|c c c c |c}
\hline
\textbf{CBSD68}&\multicolumn{12}{|c|}{\textbf{Blur Kernels}} \\
 \cline{2-13}
 \textbf{SR Methods}&\textbf{\textit{I}}& \textbf{\textit{II}} & \textbf{\textit{III}}&\textbf{\textit{IV}}&\textbf{\textit{V}} &\textbf{\textit{VI}} & \textbf{\textit{VII}} &\textbf{\textit{VIII}}&\textbf{\textit{IX}}&\textbf{\textit{X}}&\textbf{\textit{XI}}&\textbf{\textit{XII}}\\
\hline
IKR-Net  & \textbf{1.72}  & \textbf{0.65} & \textbf{0.16} & \textbf{0.06}  & \textbf{0.16}  & \textbf{0.23}  & \textbf{0.33}  & \textbf{0.05}  & \textbf{1.60}  & \textbf{1.89}  & \textbf{2.80}  & \textbf{3.10}\\
BISR-Net  & 4.03  & 3.32 & 2.43 & 1.60  & 2.08  & 3.00  & 2.74  & 1.41  & 6.83   & 6.54  & 13.4  & 5.87\\
FKP  & 17.3 & 0.95 & 0.37 & 0.37 & 0.24 & 0.64 & 0.35 & 0.29 & 8.21 & 10.4 & 12.0 & 10.8\\
\hline
\end{tabular}
\label{tab_ker}
\end{center}
\end{table*}

\begin{figure}[htbp]
\centerline{
\begin{subfigure}{0.25\textwidth}\includegraphics[scale=0.25]{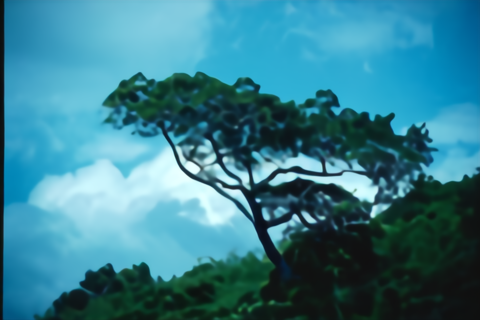}\caption{Known}\end{subfigure}
\begin{subfigure}{0.25\textwidth}\includegraphics[scale=0.25]{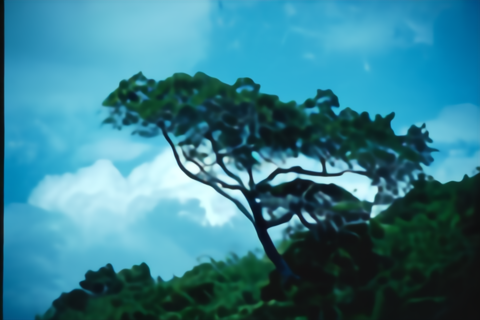}\caption{Predicted}\end{subfigure}}
\centerline{
\begin{subfigure}{0.25\textwidth}\includegraphics[scale=0.25]{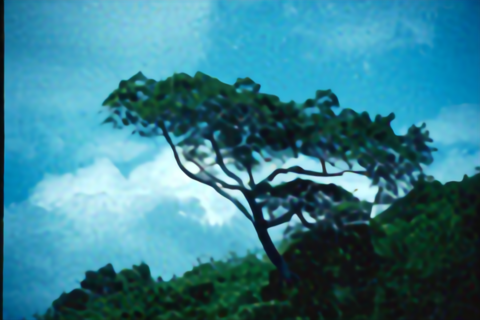}\caption{Zero}\end{subfigure}
\begin{subfigure}{0.25\textwidth}\includegraphics[scale=0.25]{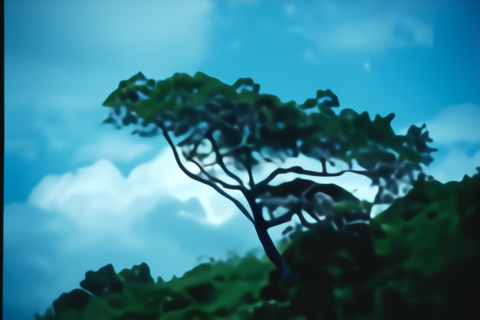}\caption{Max}\end{subfigure}}
\caption{Visual results for IKR-Net when the noise variance is known, predicted, assumed zero and assumed \%3 (scale factor $\times 4$).}
\label{abla_noise}
\end{figure}

In Table \ref{tab_known_noise}, an ablation study is provided for the noise estimation module $\mathcal{F}$. IKR-Net performance is evaluated for the following cases: i) Known: the noise variance is assumed to be known; ii) Predicted: noise variance is predicted by $\mathcal{F}$; iii) Zero: the image is assumed to have no noise;  iv) Max:  the image is assumed to have maximum noise of 3\%. Average PNSR values are provided for the three class of kernels: isotropic Gaussian, anisotropic Gaussian and motion kernels. As seen from the table, perfect knowledge of the noise variance has no effect on the average PSNR performance. The hyper-parameter estimator module  $\mathcal{F}$ provides very accurate estimation of noise standard deviation, the estimation error being typically much less than 10\% of the actual value. IKR-Net performance is also very robust against any errors in noise estimation, thereby providing almost the same performance with and without any prior knowledge of the noise variance. On the other hand, SR output quality of IKR-Net drops significantly when the noise estimation module is not used. The average loss of PSNR is about 0.5 dB  if the model assumes zero noise at the input. When maximum noise is assumed for the input image, the PSNR drop is only 0.05 dB on average. Figure  \ref{abla_noise} compares the visual results for Known, Predicted, Zero and Max test cases. It is clear from the figures that Zero scenario outputs a noisy SR image, while Max noise scenario over-smooths the model output. Despite little change in PNSR, maximum noise assumption produces blurry SR images, thereby reducing the visual quality. On the other hand, there is little visual difference between Predicted and Known case results in the figure.


\begin{table}
\caption{Comparison of IKR-Net performance with known, estimated, zero and maximum noise standard deviation (CBSD68 dataset,  2\% noise).}
\begin{center}
\begin{tabular}{|c c| c |c |c | c|}
\hline
\multicolumn{2}{|c|}{\multirow{2}{*}{\textbf{CBSD68}}}&\multicolumn{4}{|c|}{\textbf{Blur Kernels}} \\
 \cline{3-6}
 & & \textbf{iso.} & \textbf{aniso.} & \textbf{motion} &\textbf{Av.}\\
\hline
\multirow{4}{*}{$\times 4$} & \multicolumn{1}{|c|}{Predicted} & 25.31 & 25.01 & 23.58 & 24.63\\
\cline{2-6}
                            & \multicolumn{1}{|c|}{Known}     & 25.31 & 25.01 & 23.58 & 24.63 \\
                            \cline{2-6}
                            & \multicolumn{1}{|c|}{Zero}     & 24.59 & 23.87 & 23.05 & 23.84 \\ 
                            \cline{2-6}
                            & \multicolumn{1}{|c|}{Max}     & 25.24 & 24.93 & 23.58 & 24.58 \\
\hline
\hline
\multirow{4}{*}{$\times 2$} & \multicolumn{1}{|c|}{Predicted} & 27.93 & 25.97 & 24.92 & 26.27 \\
\cline{2-6}
                                & \multicolumn{1}{|c|}{Known} & 27.93 & 25.98 & 24.92 & 26.27\\
                                \cline{2-6}
                                 & \multicolumn{1}{|c|}{Zero} & 27.67 & 25.75 & 24.71 & 26.04\\
                                 \cline{2-6}
                                  & \multicolumn{1}{|c|}{Max} & 27.84 & 25.91 & 24.90 & 26.22\\
\hline
\end{tabular}
\label{tab_known_noise}
\end{center}
\end{table}

\begin{figure}[htbp]
\centerline{
\begin{subfigure}{0.25\textwidth}\includegraphics[scale=0.25]{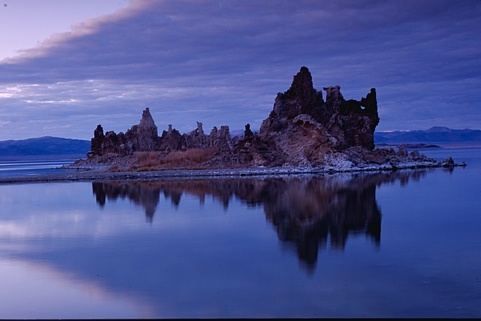}\caption{HR image}\end{subfigure}
\begin{subfigure}{0.25\textwidth}\includegraphics[scale=0.25]{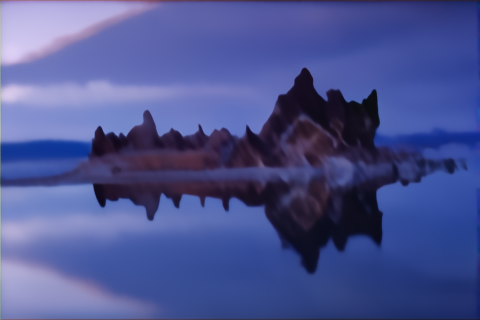}\caption{IKR with SCU-Net}\end{subfigure}}
\centerline{
\begin{subfigure}{0.25\textwidth}\includegraphics[scale=0.25]{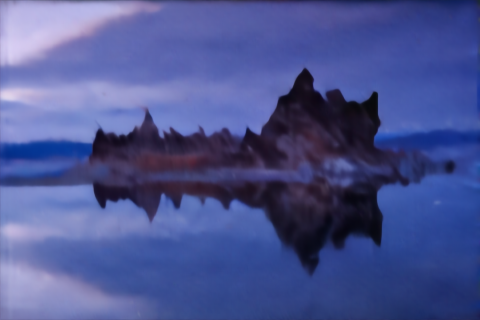}\caption{IKR+Denoise}\end{subfigure}
\begin{subfigure}{0.25\textwidth}\includegraphics[scale=0.25]{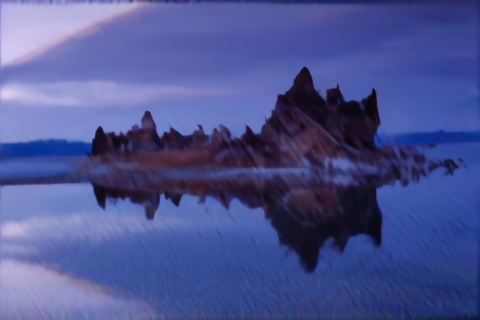}\caption{Denoise+IKR}\end{subfigure}}
\caption{Visual results for IKR with SCU-Net, Denoise+IKR, IKR+Denoise (scale factor $\times 4$,  \%3 noise).}
\label{fig_scu_ikr}
\end{figure}

Iterative application of noise estimation by module $\mathcal{F}$ and denoising by module $\mathcal{P}$ is essential for producing high quality SR image outputs. In order to analyze the contribution of iterative noise estimation and denoising on the performance of the model, pre-SR and post-SR denoising approaches are tested and compared against IKR-Net. SCU-Net color image denoising model \citep{zhang2022practical} is used for pre-SR/post-SR denoising. This model is also used in the last 4 iterations of IKR-Net as module $\mathcal{P}$ in place of ResUNet (note that, only last 4 iterations are replaced due to the heavy computational complexity of SCU-Net). Hence, the model is tested under the following three settings:
\begin{itemize}
\item Denoise+IKR: Denoise LR image, then apply IKR-Net without noise estimation
\item IKR+Denoise: Apply IKR-Net without noise estimation, then denoise SR image 
\item IKR with SCU-Net:  IKR-Net with noise estimation, where SCU-Net is used for denoising in the last 4 iterations
\end{itemize} 
 The SCU-Net model is fine-tuned separately for Denoise+IKR, IKR+Denoise and IKR-Net with SCU-Net scenarios, using LR/SR images generated with IKR-Net training set. Table \ref{scu_ikr} and Figure \ref{fig_scu_ikr} give the average PNSR and visual comparison of the three tested methods, respectively. Table \ref{scu_ikr} includes the original IKR-Net model performance for reference as well. Even though average PSNR difference is about 0.1 dB, both  Denoise+IKR and IKR+Denoise methods produce visually inferior results. While IKR+Denoise generates more blurry outputs, Denoise+IKR produces SR images with left-over noise and reconstruction artifacts. IKR with SCU-Net model has similar performance to the original IKR-Net with slightly better results for motion kernels. Hence, it can be argued that  iterative denoising in IKR-Net using well-estimated noise variance is critical for generating sharp SR images that are free of noise and reconstruction artifacts.

\begin{table}
\caption{Comparison of PSNR performance for IKR-Net, IKR with SCU-Net, Denoise+IKR, IKR+Denoise  (CBSD68 dataset, scale factor $\times 4$, 3\% noise).}
\begin{center}
\begin{tabular}{|l | c |c |c | c|}
\hline
\multicolumn{1}{|c|}{\multirow{2}{*}{\textbf{Methods}}}&\multicolumn{4}{|c|}{\textbf{Blur Kernels}} \\
\cline{2-5}
  & \textbf{iso.} & \textbf{aniso.} & \textbf{motion} &\textbf{Av.}\\
\hline
 \multicolumn{1}{|l|}{Denoise+IKR}        & 24.92 & 24.44 & 23.32 & 24.23\\
\hline

 \multicolumn{1}{|l|}{IKR+Denoise}     & 24.90 & 24.43 & 23.23 & 24.19\\
\hline

\multicolumn{1}{|l|}{IKR with SCU-Net}     & 25.00 & 24.62 & 23.35 & 24.33\\
\hline
\multicolumn{1}{|l|}{Original IKR-Net}     & 25.00 & 24.63 & 23.32 & 24.32\\
\hline
\end{tabular}
\label{scu_ikr}
\end{center}
\end{table}

Table \ref{tab_iter} compares IKR-Net performance for 16 iterative updates versus 8 iterations in scale $\times 4$. IKR-Net model is fine-tuned for optimal accuracy in 8 iterations. The average loss of PSNR is about 0.1 dB for Gaussian kernels and 0.25 dB for motion kernels. Hence the model complexity can be reduced significantly with slight loss in SR image quality.


\begin{table}
\caption{Comparison of 16 vs. 8 iterations for IKR-Net (CBSD68 dataset, no noise).}
\begin{center}
\begin{tabular}{|c c| c |c |c | c|}
\hline
\multicolumn{2}{|c|}{\multirow{2}{*}{\textbf{CBSD68}}}&\multicolumn{4}{|c|}{\textbf{Blur Kernels}} \\
 \cline{3-6}
 & & \textbf{iso.} & \textbf{aniso.} & \textbf{motion} &\textbf{Av.}\\
\hline
\multirow{2}{*}{$\times 4$} & \multicolumn{1}{|c|}{16 iter.}    & 25.88 & 25.88 & 24.36 & 25.37\\
\cline{2-6}
                            & \multicolumn{1}{|c|}{8 iter.}     & 25.77 & 25.79 & 24.11 & 25.22 \\
\hline
\end{tabular}
\label{tab_iter}
\end{center}
\end{table}

Table \ref{tab_ker_init} compares IKR-Net performance with and without using iterative kernel reconstruction (i.e. using kernel initializer only) in scales $\times 4$ and $\times 2$. The average loss of PSNR is about 0.5 dB when $\mathcal{D}k$-$\mathcal{P}k$ modules are removed. The difference in performance between the two test scenarios highlights the importance of iterative kernel refinement in order to obtain the optimal SR reconstruction quality.

\begin{figure}[htbp]
\captionsetup[subfigure]{labelformat=empty}
\centerline{
\begin{subfigure}{0.125\textwidth}\includegraphics[scale=0.3]{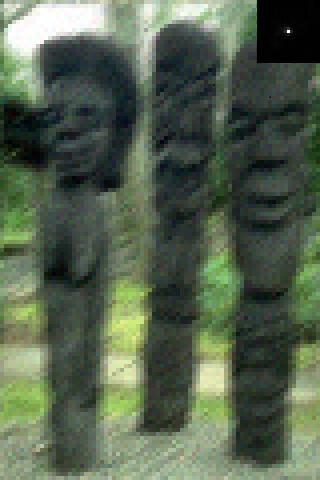}\caption{it.0}\end{subfigure}\hspace{-0.26 cm}
\begin{subfigure}{0.125\textwidth}\includegraphics[scale=0.3]{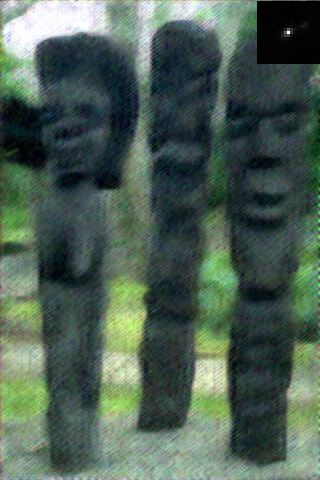}\caption{it.2}\end{subfigure}\hspace{-0.26 cm}
\begin{subfigure}{0.125\textwidth}\includegraphics[scale=0.3]{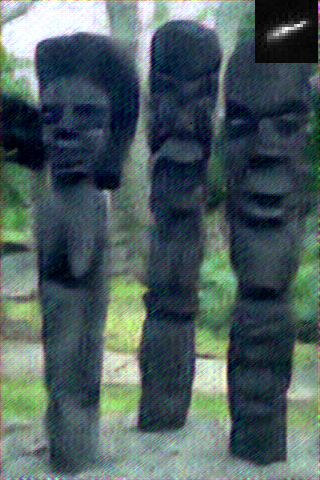}\caption{it.5}\end{subfigure}\hspace{-0.26 cm}
\begin{subfigure}{0.125\textwidth}\includegraphics[scale=0.3]{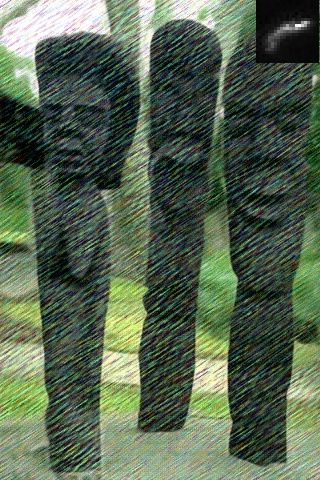}\caption{it.7}\end{subfigure}\hspace{-0.26 cm}}
\hspace{0.20 cm}\centerline{
\begin{subfigure}{0.125\textwidth}\includegraphics[scale=0.3]{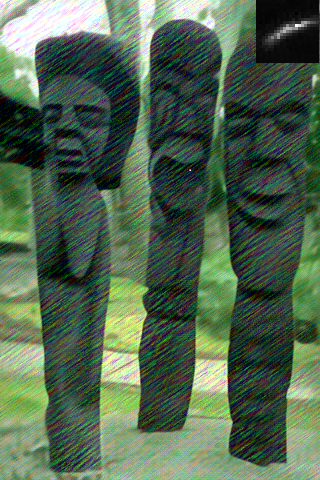}\caption{it.9}\end{subfigure}\hspace{-0.26 cm}
\begin{subfigure}{0.125\textwidth}\includegraphics[scale=0.3]{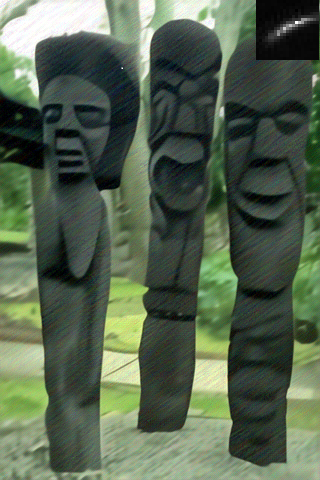}\caption{it.12}\end{subfigure}\hspace{-0.26 cm}
\begin{subfigure}{0.125\textwidth}\includegraphics[scale=0.3]{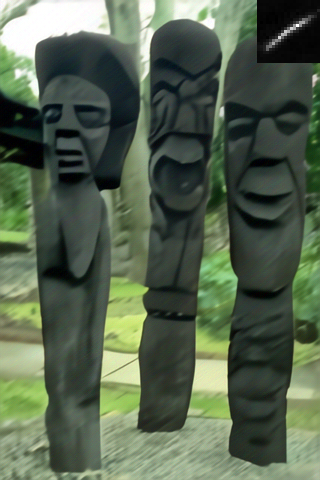}\caption{it.14}\end{subfigure}\hspace{-0.26 cm}
\begin{subfigure}{0.125\textwidth}\includegraphics[scale=0.3]{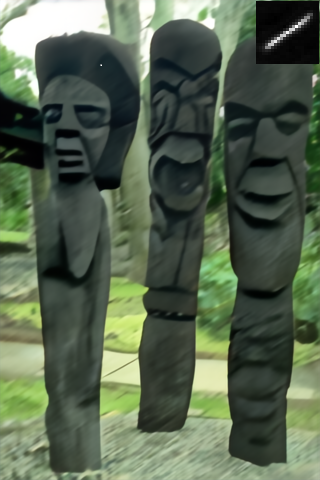}\caption{it.16}\end{subfigure}
}
\caption{SR image and blur kernel estimates at different iterations of IKR-Net (scale factor $\times 4$).}
\label{fig10}
\end{figure}

Figure \ref{fig10} shows the updated SR image and kernel estimate through different iterations of IKR-Net for a noisy LR input image. The visuals highlight the iterative denoising by the SR reconstruction module. The module $\mathcal{P}$ not only denoises the SR image but also corrects any reconstruction artifacts due to the inverse filtering in module $\mathcal{D}$. The iterative improvement of the blur kernel estimate also proves the effectiveness of $\mathcal{D}k/\mathcal{P}k$ modules for kernel reconstruction.

\begin{table}
\caption{Comparison of IKR-Net performance with and without iterative kernel reconstruction (CBSD68 dataset, no noise).}
\begin{center}
\begin{tabular}{|c c| c |c |c | c|}
\hline
\multicolumn{2}{|c|}{\multirow{2}{*}{\textbf{CBSD68}}}&\multicolumn{4}{|c|}{\textbf{Blur Kernels}} \\
 \cline{3-6}
 & & \textbf{iso.} & \textbf{aniso.} & \textbf{motion} &\textbf{Av.}\\
\hline
\multirow{2}{*}{$\times 4$} & \multicolumn{1}{|c|}{with $\mathcal{D}k$-$\mathcal{P}k$}        & 25.88 & 25.88 & 24.36 & 25.37\\
\cline{2-6}
                            & \multicolumn{1}{|c|}{without $\mathcal{D}k$-$\mathcal{P}k$}     & 25.49 & 25.39 & 23.55 & 24.81\\
\hline
\hline
\multirow{2}{*}{$\times 2$} & \multicolumn{1}{|c|}{with $\mathcal{D}k$-$\mathcal{P}k$}        & 29.63 & 27.11 & 25.57 & 27.43\\
\cline{2-6}
                                & \multicolumn{1}{|c|}{without $\mathcal{D}k$-$\mathcal{P}k$} & 29.21 & 26.58 & 25.30 & 27.03 \\
\hline
\end{tabular}
\label{tab_ker_init}
\end{center}
\end{table}

\section{Conclusion and Future Work}
This article has proposed a blind iterative SISR framework that can be trained end-to-end for joint reconstruction of the blur kernel and the SR image. The proposed architecture contains separate modules for kernel estimation, noise estimation and SR image reconstruction. Thanks to its modular architecture, the proposed method can generalize well on  blur kernels of various types including isotropic, anisotropic Gaussian and motion blurs. The noise and hyper-parameter estimation module can handle various levels of additive noise in the input image and provide parameters for optimal level of filtering for noise/artifact-free results with fine high-frequency details. Reconstruction errors due to initial kernel mismatches can be compensated via iterative refinement approach, resulting in superior performance when compared to the other state-of-the-art blind SR methods in terms of both qualitative and quantitative comparisons. 

In the future, we plan to extend this work by incorporating other types degradations (e.g. compression artifacts, camera noise, spatially varying blurs, etc.) into the proposed framework and training procedure. Also, other network topologies, such as attention mechanisms and transformers, will be tested for kernel estimation and SR image reconstruction. The proposed framework can be easily applied to other inverse problems in imaging, such as deblurring/denoising and medical image reconstruction / enhancement. 

\section*{Acknowledgments}
This work is supported in part by TUBITAK Grant Project No: 119E566.

\bibliographystyle{model2-names}
\bibliography{refs}

\end{document}